\documentclass[review,3p]{elsarticle}
\usepackage[utf8]{inputenc}

\usepackage{hyperref}
\usepackage{color, colortbl, soul}
\soulregister\cite7
\soulregister\ref7
\soulregister\pageref7
\soulregister\url7
\soulregister\textit7
\definecolor{lightyellow}{cmyk}{0,0,0.50,0}
\sethlcolor{lightyellow}
\definecolor{yellow}{cmyk}{0,0,0.50,0}
\usepackage[cmex10]{amsmath}
\usepackage{balance}
\usepackage{makeidx}
\usepackage{amssymb}
\usepackage{latexsym}
\usepackage{graphicx}
\usepackage[tight,footnotesize]{subfigure}
\usepackage{comment}
\usepackage{booktabs}
\definecolor{Gray}{gray}{0.75}
\definecolor{LGray}{gray}{0.95}

\newlength{\barwidth}
\def\mybar#1{{\color{Gray}\rule{#1\barwidth}{6.5pt}}}

\newcommand{\cut}[1]{}
\newcommand{\set}[1]{{\small \emph{#1}}}

\begin{document}
\title{Fame for sale: efficient detection of fake Twitter followers\tnoteref{t1}}
\tnotetext[t1]{An extended version of this preprint has been accepted for publication on Elsevier's journal Decision Support Systems \url{http://dx.doi.org/10.1016/j.dss.2015.09.003}. Please, cite this work as: ``Cresci, S., Di Pietro, R., Petrocchi, M., Spognardi, A., \& Tesconi, M. (2015). Fame for sale: Efficient detection of fake Twitter followers. Decision Support Systems, 80, 56-71".}

\makeatletter{}\author[iit,bell]{Stefano Cresci}
\ead{stefano.cresci@iit.cnr.it}
\author[bell]{Roberto Di Pietro}
\ead{roberto.di\_pietro@alcatel-lucent.com}
\author[iit]{Marinella Petrocchi}
\ead{marinella.petrocchi@iit.cnr.it}
\author[iit]{Angelo Spognardi\corref{cor1}\fnref{fn1}}
\ead{angelo.spognardi@iit.cnr.it}
\author[iit]{Maurizio Tesconi}
\ead{maurizio.tesconi@iit.cnr.it}
\address[iit]{IIT-CNR, Via G. Moruzzi, 1 -- 56124 Pisa, Italy}
\address[bell]{Bell Labs, Alcatel-Lucent, Paris, France}
\cortext[cor1]{Corresponding author}
\fntext[fn1]{Phone number: +39 050 315 3376} 

\begin{abstract}
  \textit{Fake followers} are those Twitter accounts specifically created to inflate
  the number of followers of a target account. 
    Fake followers are dangerous for the social platform and beyond,
  since they may alter concepts like popularity and influence in the
  Twittersphere---hence impacting on economy, politics, and society.
  In this paper, we contribute along different dimensions. First, we
  review some of the most relevant existing features and rules
  (proposed by Academia and Media) for anomalous Twitter accounts
  detection. Second, we create a baseline dataset of verified human
  and fake follower accounts.
    Such baseline dataset is publicly available to the scientific
  community.  Then, we exploit the baseline dataset to train a set of
  machine-learning classifiers built over the reviewed rules and
  features. Our results show that most of the rules proposed by Media
  provide unsatisfactory performance in revealing fake followers,
  while features proposed in the past by Academia for spam detection
  provide good results. Building on the most promising features, we
  revise the classifiers both in terms of reduction of overfitting and
  cost for gathering the data needed to compute the features. The
  final result is a novel \textit{Class A} classifier, general enough
  to thwart overfitting, lightweight thanks to the usage of the less
  costly features, and still able to correctly classify more than 95\%
  of the accounts of the original training set. We ultimately perform
  an information fusion-based sensitivity analysis, to assess the
  global sensitivity of each of the features employed by the
  classifier.
  \\*
  The findings reported in this paper, other than being supported by a
  thorough experimental methodology and interesting on their own, also
  pave the way for further investigation on the novel issue of fake Twitter followers.
  
\end{abstract}


\maketitle
\makeatletter{}
\section{Introduction}
\label{sec:intro}

Originally started as a personal microblogging site, 
Twitter has been transformed by common use to an information publishing venue. 
Statistics reported about a billion of Twitter subscribers, with 302 million monthly active users\footnote{C. Smith, By The Numbers: 150+ Amazing Twitter Statistics, http://goo.gl/o1lNi8 - June 2015. Last checked: July 23, 2015.}.
 Twitter annual advertising revenue in 2014 has been estimated to around \$480 million\footnote{Statistic Brain, Twitter statistics, http://goo.gl/XEXB1 - March 2015. Last checked: July 23, 2015.}.
 Popular public characters, such as actors and singers, as well as traditional mass media (radio, TV, and newspapers) use Twitter as a new media channel.

Such a versatility and spread of use have made Twitter the ideal arena for proliferation of 
anomalous accounts, that behave in unconventional ways. Academia has mostly focused its attention on \textit{spammers}, those accounts actively putting their efforts in spreading malware, 
sending spam, and advertising activities of doubtful legality~\cite{chu2012detecting, Yang:2013, ahmed2013generic, miller2014twitter}.
To enhance their effectiveness, these malicious accounts are often armed with automated twitting programs, as stealthy as to mimic real users, known as {\em bots}. 
In the recent past, media have started reporting that the accounts of
politicians, celebrities, and popular brands featured a suspicious
inflation of followers\footnote{Corriere Della Sera (online ed.), Academic Claims 54\% of Grillo's Twitter Followers are Bogus, http://goo.gl/qi7Hq - July 2012. Last
checked: July 23, 2015.}. So-called {\it fake followers} correspond to Twitter accounts
specifically exploited to increase the number of followers of a target
account.  As an example, during the 2012 US election campaign,
the Twitter account of challenger Romney experienced a sudden jump in
the number of followers. The great majority of them has been later
claimed to be fake\footnote{New York Times (online ed.), Buying Their Way to Twitter Fame, http://goo.gl/VLrVK, -  August
2012. Last checked: July 23, 2015.}. Similarly, before the last general Italian elections (February 2013), online blogs and newspapers had reported statistical data over a supposed percentage of fake followers of major candidates\footnote{The Telegraph (online ed.), Human or bot? Doubts over Italian comic Beppe Grillo's Twitter followers,
http://goo.gl/2yEgT - July 2012. Last checked: July 23, 2015.}.
At a first glance, acquiring fake followers could seem a practice limited to foster one's vanity---a maybe questionable, but harmless practice.
However, artificially inflating the
number of followers can also be finalized to make an account more
trustworthy and influential, in order to stand from the crowd and to
attract other genuine followers~\cite{cha2010measuring}. Recently, banks and
financial institutions in U.S. have started to analyze Twitter and Facebook accounts of
loan applicants, before actually granting the loan\footnote{Le Monde (online ed.), Dis-moi combien d'amis tu as sur Facebook, je
te dirai si ta banque va t'accorder un pr\^
et, http://goo.gl/zN3PJX - Sept. 2013. Last checked: July 23, 2015.}.\cut{Indeed, the more the supposed
influence, the more those accounts with lots of followers will likely
interfere with the genuine followers.} Thus, 
to have a
``popular" profile can definitely help to augment the creditworthiness of the applicant.  \cut{For example,
  politicians or columnists to unnaturally enlarge their interference
  on the users.}
Similarly, if the practice of buying fake followers is adopted by
malicious accounts, as spammers, it can act as a way to post more
authoritative messages and launch more effective advertising
campaigns~\cite{castillo2011information}. 
Fake followers detection seems to be an easy task for many bloggers,
that suggest their ``golden rules'' and provide a series of criteria
to be used as red flags to classify a Twitter account behavior.
However, such rules are usually paired neither with analytic
algorithms to aggregate them, nor with validation mechanisms.  As for
Academia, researchers have focused mainly on spam and bot detection,
with brilliant results characterizing Twitter accounts based on their
(non-)human features, mainly by means of machine-learning classifiers trained
over manually annotated sets of accounts.

To the best of our knowledge, however, despite fake followers constitute a widespread
phenomenon with both economical and social impacts, in the literature the topic 
has not been deeply investigated yet.

\subsubsection*{Contributions}
The goal of this work is to shed light 
on the phenomenon of fake Twitter followers, aiming at overcoming current limitations in their characterization and detection. In particular, we provide the following contributions. 
First, we build a baseline dataset of Twitter accounts where humans and fake followers are known \textit{a priori}. 
Second, we test known methodologies for bot and spam
detection on our baseline dataset. In particular, we test the Twitter accounts in our reference set against algorithms based on: (i) single classification rules proposed by bloggers, and (ii) feature sets  proposed in the literature for detecting spammers. The results of the analysis suggest that fake followers detection deserves specialized mechanisms: specifically, algorithms based on classification rules do not succeed in detecting the fake followers in our baseline dataset. Instead, classifiers based on features sets for spambot detection work quite well also for fake followers detection. Third, we classify all the investigated rules and features based on the cost required for gathering the data needed to compute them. Building on theoretical calculations and empirical evaluations, we show how the best performing features are also the most costly ones. The novel results of our analysis show that data acquisition cost often poses a serious limitation to the practical applicability of such features.
Finally, building on the crawling cost analysis, we design and
implement lightweight classifiers that make use of the less costly
features, while still being able to correctly classify more than 95\%
of the accounts of our training dataset. In addition, we also
  validated the detection performances of our classifiers over two
  other sets of human and fake follower accounts, disjoint from the
  original training dataset.

\subsubsection*{Road map}The remainder of this paper is structured as follows.
Section~\ref{sec:RW} considers and compares related work in the area of Twitter
spam and bot detection. Section~\ref{sec:datasets} describes our baseline dataset.
In Section~\ref{sec:fakedet}, we evaluated a set of criteria for
fake Twitter followers detection promoted by Social Media analysts using our baseline dataset.
In Section~\ref{sec:spamdet}, we examine features used in previous works
for spam detection of Twitter accounts.  In Section~\ref{sec:opt-cost} we compute the cost for extracting the
features our classifiers are based on. A lightweight and efficient
classifier is also provided, attaining a good balance between fake
followers detection capability and crawling cost.     Finally, Section~\ref{sec:conc} concludes the paper.

\makeatletter{}
\section{Related Work}\label{sec:RW}

Quoting from~\cite{asonam14}, ``A fake Twitter account is
  considered as one form of deception (i.e., deception in both the
  content and the personal information of the profiles as well as
  deception in having the profile follow others not because of
  personal interest but because they get paid to do so)."  
  The second characterization for deception is exactly the one we deal with in
  our paper. We specifically consider \textit{fake followers} as those
  Twitter accounts appropriately created and sold to customers, which
  aim at magnifying their influence and engagement to the eyes of the
  world, with the illusion of a big number of followers.

So defined fake followers are only an example of anomalous
  accounts which are spreading over Twitter. Anomalies have been
  indeed identified in the literature as either spammers
  (i.e. accounts that advertise unsolicited and often harmful content,
  containing links to malicious pages~\cite{Stringhini:2010}), or bots
  (i.e., computer programs that control social accounts, as stealthy
  as to mimic real users~\cite{SocialBot11}), or cyborgs (i.e.,
  accounts that interweave characteristics of both manual and
  automated behavior~\cite{ChuGWJ:2012}). Finally, there are fake
  followers, accounts massively created to follow a target
  account and that can be bought from online accounts markets. 
\subsection{Grey literature and Online Blogs}
Before covering the academic literature, we briefly report on online
documentation that presents a series of intuitive fake follower
detection criteria, though not proved to be effective in a scientific
way. The reason why we cite this work is twofold: on the one hand,
online articles and posts testify the quest for a correct
discrimination between genuine and fake Twitter followers; on the
other hand, we aim at assessing in a scientific manner whether such criteria could actually be employed for fake followers detection.

As an example, a well-known blogger in~\cite{blogger1:2012} indicates
as possible bots-like distinctive signals the fact that bots accounts:
1) have usually a huge amount of following and a small amount of
followers; 2) tweet the same thing to everybody; and, 3) play the
follow/unfollow game, i.e. they follow and then unfollow an account
usually within 24 hours.  Criteria advertised by online blogs are
mainly based on common sense and the authors usually do not even
suggest how to validate them.

A series of reports published by the firm
\textit{Digital evaluations} \cite{Camisani:2012} have attracted the attention of Italian and European newspapers and
magazines, raising doubts on the Twitter popularity of politicians and
leading international companies. A number of criteria, inspired by
common sense and denoting \textit{human}
behavior, are listed in the reports and used to evaluate a sample of the
followers of selected accounts. For each criterion satisfied by a
follower, a \textit{human} score is assigned.  For each not fulfilled
criterion, either a \textit{bot} or \textit{neutral} score is assigned. According to the total score, Twitter
followers are classified either as humans, as bots or as neutral (in
the latter case, there is not enough information to assess their
nature), providing a quality score of the effective influence of the
followed account.  The results in~\cite{Camisani:2012}, however, lack
a validation phase.

Finally, some companies specialized in social media analysis offer online
services to estimate how much a Twitter account is \textit{genuine} in
terms of its followers
~\cite{StatusPeople,SocialBakers,TwitterAudit}.
However, the criteria used for the analysis
are not publicly disclosed and just partially deductible from
information available on their web sites. Moreover, as demonstrated in
our previous work~\cite{DASec:2014}, these analyses are affected by several
biases like small and statistically unsound sampling strategies.

\subsection{Academic literature}
In recent years, spam detection on Twitter has been the matter of
  many investigations, approaching the issue from several points of
  view. As an example, a branch of research focused on the textual
  content of tweets~\cite{miller2014twitter, ahmed2013generic, GaoCLPC12},
  studying the redirection of embedded URLs in tweets~\cite{Lee:2013}, and classifying the URLs landing pages~\cite{ThomasGMPS11}.
Other works tackled the problem of deception on Twitter via epistemology. For instance, in~\cite{zubiaga2014tweet} authors evaluate 4 epistemological features for the task of deception detection: authority, plausibility and support, independent
corroboration, and presentation.

  Work in~\cite{Gao14} overcomes the limitation of not being able to correctly label those tweets without URLs as spam tweets,
  by proposing a composite tool, able to match incoming tweets with
  underlying templates commonly used by spammers.  
Instead of considering the content of tweets, work
  in~\cite{asonam14} tries to classify if an account can be trusted or not
   based on possibly inconsistent information originating from the
  profile of the account only.

A series of works investigate spammers on microblogging platforms
  through a multi-feature approach, including features on the profile,
  the behavior, and the timeline of an account.  Within this
research line, we cite here~\cite{Stringhini:2010},~\cite{Yang:2013},
and~\cite{weibo14}.  The work in~\cite{Stringhini:2010} presents an
analysis on how spammers operate on Facebook, Twitter, and
MySpace, reporting that the suspicious accounts shared some
common traits on specific features.
Those served as input to a machine learning-based
classifier~\cite{data_mining} leading to 
the detection of more than 15,000 spam profiles, that Twitter promptly
deleted.
In~\cite{Yang:2013}, the authors propose a taxonomy of criteria for
detecting Twitter spammers. A series of experiments show how the newly
designed criteria have higher detection rates, when compared to the
existing
ones. In~\cite{weibo14}, the authors leverage a combination of
  behavioral features (such as tweeting and retweeting activities),
  network features (such as the number of an account's followers and
  friends), and content-based features to develop a hybrid
  mathematical model for spammer detection in Weibo, the Chinese
  microblogging site resembling Twitter.

The authors of~\cite{ChuGWJ:2012} classify Twitter accounts in three
classes: humans, bot, and cyborgs. The latter class represents either
bot-assisted humans or human-assisted bots. The authors used a decision maker
based on three specialized components that considered the content and timing of tweets and some account properties.

The algorithms proposed in~\cite{ComCom14, miller2014twitter} aim at spotting groups of
automated malicious Twitter accounts as quickly as possible, to avoid
the accounts' owners from taking advantage of them.Thus, authors apply clustering algorithms to group accounts created within a short period of time, considering, among others, name- and content-based features.

\label{subset:markets} 
In~\cite{Stringhini:2012}, the authors list several criteria to detect
clients and victims of Twitter account markets, that are online
services offering their subscribers to provide followers in exchange
for a fee, and to spread promotional tweets on their behalf. In
another work~\cite{Stringhini:2013}, the same research team provides
more details about the account markets, analyzing additional properties
and characteristics of their customers (e.g., the dynamics of
followers and friends and the ability of generating engagement), and
providing a classifier for the detection of both markets and market
customers. 
The authors of~\cite{Thomas2013} monitor prices, availability, and
  fraud perpetrated by a set of merchants of Twitter accounts over the
  course of a ten-months period.  Such a research is a spotlight on
  techniques and methodologies that accounts markets exploit to create
  and register fraudulent accounts, from CAPTCHA solving services, to
  deceitful email credentials and a diverse pool of IP addresses to
  evade blacklisting. In collaboration with Twitter itself, the
  authors developed a classifier to detect such fraudulent accounts,
  which were consequently suspended. 
 
It is worth noting that the cited selection of academic works is not
  exhaustive. However, it considers a huge collection of criteria,
  which we further leverage for our approach to fake Twitter followers
  detection.  There are other works for spam detection, not detailed here, 
  like~\cite{chu2012detecting,Lee:2010,Yardi,yan2013peri,mccord:2011, bhat2014using}, which base their results on subsets, or on
  slightly modified versions, of criteria considered by the
  selected set of related work.

\subsection{Differences and similarities with our approach}
The goal of our research is the automatic detection of those
  Twitter accounts specifically created to inflate the number of
  followers of some target account (so called fake Twitter
  followers). A priori, both spammers, bots, and genuine users'
  accounts could fall in the macro-category of fake followers, and
  specific features already proved effective in the literature for
  spotting spammers and bots could work also in the case of fake
  followers. It was indeed this observation that initially drove the
  authors of this paper towards the direction of testing rules and
  features from past works on a reference dataset of genuine accounts
  and fake followers. This contributed to prune those rules and
  features that behaved worst in detecting fake followers, and leave
  the ones that well behave.

From a technical point of view, in our experiments we rely on
  machine learning-based classifiers exploiting features of 1)
  profile, 2) activity, and 3) relationships of the accounts, similarly to~\cite{Stringhini:2010,Yang:2013}.
  Instead, we do not rely on features inherent to specific contents of tweets, such as the
  presence of URLs and the semantics of the text~\cite{GaoCLPC12,ThomasGMPS11}. 
We move beyond the mere application of already tested features to
  a new dataset, since we revise our classifiers to reduce overfitting
  and cost for data gathering, as illustrated in
  Sections~\ref{sec:spamdet} and~\ref{sec:opt-cost}.  

Finally, similarly to~\cite{Stringhini:2013}, we
  bought fake Twitter followers from different markets available on
  the Web. We conducted such an exercise independently
  from~\cite{Stringhini:2013} and, moreover, goals
  of the two works are quite different, ours focusing on
  accounts sold by these markets, while the other targeting their
  customers.  As for the genuine part of our baseline dataset, we
  \textit{recruit} accounts of people that have voluntarily adhered to
  our campaign, and leverage a dataset of annotated accounts, belonging to
  people active on Twitter within a particular period of time on a
  specific domain, and whose authenticity has been verified. However, willing to test our
  classifiers over a representative sample  of the entire Twitter population, we also
  approach the construction of a test set by randomly picking: 1) a
  sample of Barack Obama's followers, and 2) a sample of the Twitter
  population.

\makeatletter{}
\section{Baseline datasets}\label{sec:datasets}
In this section we present the datasets of Twitter accounts used
to conduct our empirical study throughout the
paper. We detail how we collected each of them and how we verified if
they were genuine humans or fake followers. Despite the final size of
the baseline dataset, to perform our research, we altogether crawled 9
millions of Twitter accounts and about 3 millions of tweets.
To foster investigation on the novel issue of fake Twitter followers, our baseline dataset has been made publicly available for research purposes~\cite{fakeproject}.

\subsection{The Fake Project}
\label{subsec:FP}

The Fake Project started its activities on December 12, 2012, with the
creation of the Twitter account \textit{@TheFakeProject}. Its profile
reports the following motto: ``{\it Follow me only if you are NOT a
  fake}" and explains that the initiative is linked with a research
project owned by researchers at IIT-CNR, in Pisa-Italy.
In a first phase, the owners contacted further researchers and journalists
to advertise the initiative and also foreign journalists and bloggers supported the initiative in their countries.
In a twelve days period (Dec 12-24, 2012), the account has been followed by 574
followers. Through the Twitter APIs, we crawled a series of
public information about these followers together with that of their followers and
followings. For this dataset, we crawled the 574 accounts, leading to
the collection of 616,193 tweets and 971,649 relationships (namely,
linked Twitter accounts).

All those followers voluntarily joined the project. To include them in
our reference set of humans, we also launched a verification phase.
Each follower received a direct message on Twitter from
\textit{@TheFakeProject}, containing an URL to a
CAPTCHA, unique for each follower.
We consider as ``certified humans'' all the 469 accounts out of the 574
followers that successfully completed the CAPTCHA.
In the remainder of this section this dataset is referred to as
{\small \emph{TFP}}.

\subsection{\#elezioni2013 dataset}\label{subsec:politics}
The \#elezioni2013 dataset, henceforth {\small \emph{E13}}, was born
to support a research initiative for a sociological study carried out
in collaboration with the University of Perugia and the Sapienza
University of Rome.  The study focused on the strategic changes in the
Italian political panorama for the 3-year period 2013-2015.
Researchers identified 84,033 unique Twitter accounts that used the
hashtag \#elezioni2013 in their tweets, during the period between
January 9 and February 28, 2013. Identification of these accounts has
been based on specific keyword-driven queries on the username and
biography fields of the accounts' profiles.  Keywords include blogger,
journalist, social media strategist/analyst, and congressperson.
Specific names of political parties have been also searched. In
conclusion, all the accounts belonging to politicians and candidates,
parties, journalists, bloggers, specific associations and groups, and
whoever somehow was officially involved in politics, have been
discarded.  The remaining accounts (about 40k) have been classified as
{\it citizens}.  This last set has been sampled (with confidence level
95\% and confidence interval 2.5), leading to a final set of 1488
accounts, that have been subject to a manual verification to determine
the nature of their profiles and tweets.
The manual verification process has been carried out by two
sociologists from the University of Perugia, Italy. It involved the
analysis of profile pictures, biographies, and timeline of the
accounts under investigation. Accounts not having a biography or a
profile picture have been discarded. URLs in biographies have also
been manually checked to allow for a deeper analysis of the
subject. Only accounts labeled as humans by both the sociologists have
been included in the {\small \emph{E13}} dataset.  Overall, the manual
verification phase lasted roughly two months. As a result, 1481
accounts became part of dataset {\small \emph{E13}}.

\subsection{Baseline dataset of human accounts}\label{subsec:humanset}
The above introduced datasets form our final set, labeled {\small
  \emph{HUM}}, of 1950 verified human accounts.  It is worth noting
how the two subsets differ from each other.  The {\small \emph{TFP}}
set consists of accounts that have been recruited on a volunteer base:
people involved in the initiative aimed to be part of an academic
study for discovering fake followers on Twitter, and are a mixture of
researchers and social media experts and journalists, mostly from
Italy, but also from US and other European countries.  The {\small
  \emph{E13}} set consists of particularly active Italian Twitter
users, with different professional background and belonging to diverse
social classes, sharing a common interest for politics, but that do
not belong to the following categories: politicians, parties,
journalists, bloggers.  
\subsection{Baseline dataset of fake followers}\label{subsec:fakeset}
In April, 2013, we bought 3000 fake accounts from three different
Twitter online markets. In particular, we bought 1000 fake accounts
from \url{http://fastfollowerz.com}, 1000 from \url{http://intertwitter.com},
and 1000 fake accounts from \url{http://twittertechnology.com}, at a price
of \$19, \$14 and \$13 respectively.
Surprisingly, \textit{fastfollowerz} and \textit{intertwitter} gave us
a few more accounts than what we paid for, respectively 1169 and 1337
instead of 1000.  We crawled all these accounts to build a
\textit{fastfollowerz} dataset, labeled {\small \emph{FSF}}, and an
\textit{intertwitter} dataset labeled {\small \emph{INT}}.  Instead,
we were unable to crawl all the 1000 fake followers bought from
\textit{twittertechnology} since 155 of them got suspended almost
immediately. The remaining 845 accounts constitute the
\textit{twittertechnology} dataset, which is labeled {\small
  \emph{TWT}}.

We acknowledge that our fake followers dataset is just  illustrative,
and not exhaustive, of all the possible existing sets of fake
followers. However, it is worth noting that we found the Twitter
accounts marketplaces by simply Web searching them on the most common
search engines. Thus, we can argue that our dataset represents what
was easily possible to find on the Web at the time of searching.

\subsection{Baseline dataset of fake followers and human
  accounts}\label{subsec:baseset}
The final baseline dataset exploited in our experiments is composed of
both fake and human profiles. 
In the following, we briefly discuss the distribution between fake and human accounts that has been chosen for this dataset. 
Many machine-learning techniques
are affected by the imbalance of
  the natural distributions of the minority and majority classes. This
  is why, for example, works in the literature have studied how the
  decision tree-based techniques perform when varying the distribution
  of the training set. In particular, Weiss and Provost, in~\cite{weiss2003learning}, 
  considered the performances of
  decision-tree based classifiers to predict the samples of 26
  different datasets, with different distributions between the
  minority and majority classes. The conclusions of their
  investigation have shown that the metric used to evaluate the
  performance of the different classifiers changes the optimal
  distribution of the classes for the training set. For example, after
  their empirical analysis, using accuracy as evaluation metric, 9 out
  of the 26 datasets have the optimal distribution very different from
  the natural one, while, when using the \textit{AUC} as evaluation
  metric, this number grows to 14 out of the 26. Moreover, the optimal
  distribution has an oversampling of the minority class (there are
  also cases when the best classifier is obtained with an oversampling
  up to 90\% of the minority class samples).
          
Here, we face another fundamental issue:
  we do not precisely know the real (natural) distribution of fake
  followers and human accounts. In 2013, the Twitter staff conjectured that the
  number of ``false or spam accounts should represent less than 5\% of
  our MAUs'' (where MAUs refer to monthly active users)~\cite{twitters1form}. However, MAUs can be assimilated neither to a random sample of Twitter
  accounts nor to the followers of a given account. Moreover,   if an account has bought  fake followers, then its distribution of fake
  followers and human followers can vary dramatically from the natural
  distribution that one can find, either among MAUs, or among all the
  Twitter accounts in the Twittersphere.  In conclusion, the estimation of 5\% as false or spam accounts, in
  the whole Twitter, can not be directly extended to the fake followers of a
  given account.
  
  \begin{table}[t]
	\footnotesize
	\centering
	\begin{tabular}{ lrrrrr }
\toprule
&&& \multicolumn{3}{c}{{\textbf{relationships}}}\\
		\cmidrule{4-6}
		\textbf{dataset} & \textbf{accounts} & \textbf{tweets} & followers & friends & total\\
		\midrule
		{\small \emph{TFP}} (@TheFakeProject)				& 469	& 563,693		& 258,494		& 241,710		& 500,204\\
		{\small \emph{E13}} (\#elezioni2013)					& 1481	& 2,068,037	& 1,526,944	& 667,225		& 2,194,169\\
		{\small \emph{FSF}} (\textit{fastfollowerz})		& 1169	& 22,910		& 11,893		& 253,026		& 264,919\\
		{\small \emph{INT}} (\textit{intertwitter})			& 1337	& 58,925		& 23,173		& 517,485		& 540,658\\
		{\small \emph{TWT}} (\textit{twittertechnology})	& 845	& 114,192		& 28,588		& 729,839		& 758,427\\
		\midrule
		{\small \emph{HUM}} (human dataset)				& 1950	& 2,631,730	& 1,785,438	& 908,935		& 2,694,373\\
		{\small \emph{FAK}} (fake dataset)					& 1950	& 118,327		& 34,553		& 879,580		& 914,133\\
		\midrule
		{\small \emph{BAS}} (baseline dataset:  {\small \emph{HUM}} $\cup$ {\small  \emph{FAK}})				& 3900	& 2,750,057	& 1,819,991	& 1,788,515	& 3,608,506\\
		\bottomrule
	\end{tabular}
	\caption{Statistics about total collected data~\cite{fakeproject}.
	\label{tab:datastat}}
\end{table}  
Although Twitter has never disclosed the total number of registered users, unofficial sources claim that the Twitter accounts created up to date are many more than the MAUs.
This is why we made a conservative assumption, considering a balanced
distribution of fake followers and human followers as our baseline
dataset. To validate this assumption, we performed the experiments in~\cite{weiss2003learning} to our dataset. 
We
progressively varied the class distribution of fake followers and
human followers in the dataset, from 5\%--95\% to 95\%--5\%
(respectively 100 humans--1900 fake followers, 1900 humans--100 fake
followers), and used the obtained dataset to train J48 classifiers,
considering their performances with cross-validation.  The trained
classifiers obtained their best results on a balanced distribution of
humans and fake followers. 
To obtain a balanced dataset, we randomly undersampled the total set
of fake accounts (i.e., 3351) to match the size of the {\small
  \emph{HUM}} dataset of verified human accounts. Thus, we built a
baseline dataset of 1950 fake followers, labeled {\small \emph{FAK}}.
The final baseline dataset for this work includes both the {\small
  \emph{HUM}} dataset and the {\small \emph{FAK}} dataset for a total
of 3900 Twitter accounts. This balanced dataset is labeled {\small
  \emph{BAS}} in the remainder of the paper and has been exploited for
all the experiments described in this work (where not otherwise
specified). Table~\ref{tab:datastat} shows the number of accounts,
tweets and relationships contained in the datasets described in this
section.

\makeatletter{}\section{Fake detection with algorithms based on classification rules}
\label{sec:criteria}
\label{sec:fakedet}
In this section, we detail three procedures, originally proposed
  by bloggers and social media analysts, explicitly conceived for fake
  followers and bot accounts detection. These proposals were introduced in~\cite{Camisani:2012, blogger1:2012, SocialBakers}. The work we
  focus on in this section is not directly attributable to 
  academic work. However, it is an example of the spreading interest
  on the phenomenon of fake Twitter followers by Media and Social
  Marketing companies.
  Although we do not expect these proposals to satisfactorily perform
  for the complex task of fake followers detection, we believe a
  thorough analysis of the proposed criteria could still provide some
  useful insights. Coincidentally, all the procedures are proposed as
  algorithms relying on a list of rules, or criteria: each account to
  be classified is checked against all the rules and the outputs of
  the checks must be combined together in order to obtain the final
  classification. Unfortunately in many cases, details on how to combine the criteria to obtain the final classification of an account is not publicly available. Details on how
  aggregation has been performed are provided in~\cite{Camisani:2012}
  only.  Driven by the provided details, we implement the full
  algorithm described in~\cite{Camisani:2012} and we present its
  detection performances in Section~\ref{sec:ecc}.  In addition, for
  each of the procedures, we report the criteria as indicated by the
  original sources and we further specify how we have implemented them
  into rules suitable to be applied over our datasets. We also detail
  the reasons for our implementation choices.

In this section, we mainly focus on the application of each single
  rule over our dataset to assess its strength (or weakness) in
  discriminating fake followers.  In
  Section~\ref{sec:opt-cost}, we combine all the rules together with
  the features analyzed in Section~\ref{sec:spamdet}, to
  assess their collective classification power. This is because a
  single rule -- or feature -- alone may not perform well in
  classifying fake and human accounts, but it may improve the
  detection if used in combination with other ones. Indeed, it is worth
  noting that some of the criteria analyzed in this section have been
  actually exploited by the classifiers built in
  Section~\ref{sec:opt-cost}.

Throughout the sequel of the paper we use the term ``friends'' to
denote the users followed by an account (i.e., if $A$ follows $B$, $B$
is a friend of $A$).

\subsection{Followers of political candidates}\label{sec:Camisani}
Camisani-Calzolari~\cite{Camisani:2012} carried out a series of tests over
samples of Twitter followers of Romney and Obama, for the last US
presidential elections, as well as for popular Italian
politicians.
In~\cite{Camisani:2012} it is detailed an algorithm to evaluate an
account based on some of its public features. The cited
algorithm has enough details to be reproducible: it assigns
\textit{human/active} and \textit{bot/inactive} scores and classifies
an account considering the gap between the sum of the two scores. In
particular, the algorithm assigns to the examined accounts 1 (or more,
where specified) \textit{human} point for each of the criteria in
Table~\ref{CCruleset}.
\begin{table}[t!]
  \begin{small}
    \begin{tabular}{p{0.47\textwidth}p{0.53\textwidth}}
\toprule
\textbf{1.}   the profile contains a name;                                  & \textbf{12.}  it has used a hashtag in at least one tweet;                                                                                                        \\
\textbf{2.}   the profile contains an image;                                & \textbf{13.}  it has logged into Twitter using an iPhone\label{mcc1};                                                                                             \\
\textbf{3.}   the profile contains a physical address;                      & \textbf{14.}  it has logged into Twitter using an Android device\label{mcc2};                                                                                     \\
\textbf{4.}   the profile contains a biography;                             & \textbf{15.}  it is connected with Foursquare\label{mcc3};                                                                                                        \\
\textbf{5.}   the account has at least 30 followers;                        & \textbf{16.}  it is connected with Instagram\label{mcc4};                                                                                                         \\
\textbf{6.}   it has been inserted in a list by other Twitter users;        & \textbf{17.}  it has logged into \textit{twitter.com} website\label{mcc5};                                                                                        \\
\textbf{7.}   it has written at least 50 tweets;                            & \textbf{18.}  it has written the userID of another user in at least one tweet, that is it posted a \textit{@reply} or a \textit{mention};                         \\
\textbf{8.}   the account has been geo-localized\label{mcc10};              & \textbf{19.}  (2*number followers) $\ge$ (number of friends);                                                                                                     \\
\textbf{9.}   the profile contains a URL;                                   & \textbf{20.}  it publishes content which does not just contain URLs;                                                                                              \\
\textbf{10.}  it has been included in another user's favorites\label{mcc8}; & \textbf{21.}  at least one of its tweets has been \textit{retwitted} by other accounts (worth 2 points)\label{mcc6};                                              \\
\textbf{11.}  it writes tweets that have punctuation\label{mcc9};           & \textbf{22.}  it has logged into Twitter through different clients (worth 3 points)\label{mcc7}.                                                                  \\
\bottomrule
    \end{tabular}
  \end{small}
\caption{Camisani-Calzolari rule set.\label{CCruleset}}
\end{table}
Moreover, the account receives 2 \textit{bot} points if it only uses
APIs. Finally, for each criterion that fails to be verified, the
account receives 1 \textit{bot} point, with the exception of criteria
8, 13, 14, 15, 16 and 17: in this cases, no \textit{bot} points are assigned. To
verify those rules, we referred to the \textit{source} metadata of the
tweets, that contains a different value representing the
platform used to post a tweet. In particular, concerning the above
rules, we considered the \textit{source} metadata with the values
\textit{iphone, android, foursquare, instagram} and \textit{web},
respectively, and we assigned 1 \textit{human} point for each of the
values found at least once within the collected tweets of the
account. For the criterion 21, 2 \textit{bot} points are
assigned if no tweets of the account have been retweeted by other
users.
Considering rule~8, {\it geo-localization} is related to
tweets.  Consequently, we set this rule as satisfied when at least
one tweet of the account has been geo-localized.
For the rule~11, {\it punctuation} has been searched in both
the profile biography and in its timeline.

\subsection{Stateofsearch.com}\label{app:blogs}
Among the several bloggers that propose their golden rules to identify
suspicious Twitter accounts, we consider the \textit{``7 signals to
  look out for recognizing Twitter bots''}, according to the founder
of the social media website
\textit{stateofsearch.com}~\cite{blogger1:2012}.  The ``7 signals to
look out for'' to recognize Twitter bots are listed in Table~\ref{stateofsearchruleset}.

\begin{table}[h!]
  \begin{small}
    \begin{tabular}{p{0.47\textwidth}p{0.53\textwidth}}
	\toprule
	\textbf{1.}   the biography of the profile clearly specifies that it is a bot account\label{blog1};	& \textbf{5.}  accounts that tweet from API are suspicious\label{blog7}; \\
	\textbf{2.}   the $\frac{\textit{friends}}{\textit{followers}}$ ratio is in the order of 100:1;			& \textbf{6.}  the response time (follow $\rightarrow$ reply) to tweets of other accounts is within milliseconds\label{blog2}; \\
	\textbf{3.}   the account tweets the same sentence to many other accounts\label{blog4};		& \textbf{7.}  the account tends to follow $\rightarrow$ unfollow other accounts within a temporal arc of 24 hours\label{blog5}. \\
	\textbf{4.}   different accounts with duplicate profile pictures are suspicious\label{blog6};		& \\
	\bottomrule
    \end{tabular}
  \end{small}
\caption{Stateofsearch.com rule set.
\label{stateofsearchruleset}}
\end{table}

The rule 3 has been implemented considering the tweet as a
single unit. We consider the last 20 tweets of each timeline.
For the rule 4, we consider the existence of a duplicate profile picture when
at least 3 accounts within the dataset have the same profile picture.
For  the rule 5, we consider as tweets posted from API all
those tweets not being posted from the website \textit{twitter.com}.
For rules 6 and 7, when looking for an
  account's friends or followers list, Twitter only gives information
  about the current list, with no details about past friends or
  followers.  Moreover, Twitter does not disclose any temporal data
  related to the moment a user stared following, or got followed by,
  another user. This means that the only way to check a user's
  follow/unfollow behavior (rule 7) is to continuously
  monitor full friends and followers complete lists. The same applies
  with respect to the measurement of the delay experienced when  a user
  follows (and replies to) other users (rule 6). As further
  detailed in Section~\ref{sec:opt-cost}, the Twitter rate limits in
  the use of the APIs makes it practically infeasible to monitor
  friends and followers lists of even a small group of users.
  Therefore, we did not apply rules 6 and 7 to our
  datasets, since that would require to continually monitor those
  accounts. This also means that those rules cannot be used to support
  an automatic detection process, since they require an interactive
  process to be evaluated.

\subsection{Socialbakers'
  FakeFollowerCheck}\label{app:online-analyzers}
\label{sec:online-analyzers}
Several companies provide online tools to classify Twitter followers
based on their {\it fakeness} degree.  Here, we consider the
``FakeFollowerCheck tool'', by Socialbakers~\cite{SocialBakers}. While
the company website provides eight criteria to evaluate the fakeness
degree of the followers of a certain account, it omits details on how
to combine such criteria to classify the accounts.  We contacted their
customer service, but we were answered that ``how the respective
criteria are measured is rather an internal information''.
The FakeFollowerCheck tool analyzes the followers of an account and
considers them likely fake when the criteria listed in Table~\ref{FakeFollowerCheckruleset} are satisfied.

\begin{table}[h!]
  \begin{small}
    \begin{tabular}{p{0.47\textwidth}p{0.53\textwidth}}
	\toprule
	\textbf{1.}   the ratio $\frac{\textit{friends}}{\textit{followers}}$ of the account under investigation is 50:1, or more;	& \textbf{5.}  more than 90\% of the account tweets are links; \\
	\textbf{2.}   more than 30\% of all the tweets of the account use spam phrases, such as ``diet'', ``make money'' and ``work from home''\label{ffc2}; & \textbf{6.}  the account has never tweeted; \\
	\textbf{3.}   the same tweets are repeated more than three times, even when posted to different accounts;		& \textbf{7.}  the account is more than two months old and still has a default profile image; \\
	\textbf{4.}   more than 90\% of the account tweets are retweets;		& \textbf{8.}  the user did not fill in neither bio nor location and, at the same time, he/she is following more than 100 accounts. \\
	\bottomrule
    \end{tabular}
  \end{small}
\caption{Socialbakers rule set.
\label{FakeFollowerCheckruleset}}
\end{table}

For the rule 2, we consider as spam phrases expressions like
``diet'' or ``make money'' or ``work from home'' (both English and
Italian translations), as suggested by the website of Socialbakers.

\subsection{Evaluation methodology}
\label{sec:evaluation}\label{sec:results}
All the criteria detailed above have been applied to the 2 verified
human accounts datasets ({\small \emph{TFP}} and {\small \emph{E13}})
as well as to all the 3351 fake followers accounts bought from the
Twitter account markets ({\small \emph{FSF}} $\cup$ {\small \emph{INT}} $\cup$
{\small \emph{TWT}}), as described in Section~\ref{sec:datasets}.

We conducted one experiment for each rule, considering two classes of
accounts, the fake followers and the human ones.
To summarize the outcomes of each experiment, we considered some
evaluation metrics based on four standard indicators, namely:
\begin{enumerate}
\item [$\bullet$]  {\it True Positive (TP)}: the number of those fake followers
  recognized by the rule as fake followers;
\item [$\bullet$]  {\it True Negative (TN)}: the number of those human followers
  recognized by the rule as human followers;
\item [$\bullet$]  {\it False Positive (FP)}: the number of those human followers
  recognized by the rule as fake followers;
\item [$\bullet$] {\it False Negative (FN)}: the number of those fake followers
  recognized by the rule as human followers.
\end{enumerate}

The meaning of each indicator is graphically highlighted by the
matrix  in Table~\ref{tab:confmatrix} (called the \textit{confusion
  matrix}), where each column represents the instances
in the predicted class, while each row represents the instances in the
actual class~\cite{kohavi98}: \definecolor{Gray}{gray}{0.9}
\begin{table}[h!]
\begin{center}
  \footnotesize
    \begin{tabular}{ccc}
      \cellcolor{white} &\multicolumn{2}{c}{\textbf{predicted class}}\\
      \cmidrule{2-3}
      \textbf{actual class} & \textit{human} & \textit{fake}\\
      \midrule
      \textit{human} & \cellcolor{Gray}\textit{TN} & \cellcolor{Gray}\textit{FP}\\
      \textit{fake}  & \cellcolor{Gray}\textit{FN} & \cellcolor{Gray}\textit{TP}\\ 
      \midrule
    \end{tabular}
\end{center}
\caption{Confusion matrix.
\label{tab:confmatrix}}
\end{table}
In order to evaluate the application of each single rule to the
accounts in the baseline dataset, we consider the following standard
evaluation metrics:
\begin{enumerate}
\item [$\bullet$]\textit{Accuracy}: the proportion of predicted true results
  (both true positives and true negatives) in the population, that is
  $\frac{\textit{TP}+\textit{TN}}{\textit{TP+TN+FP+FN}}$;
\item [$\bullet$]\textit{Precision}: the proportion of predicted positive cases
  that are indeed real positive, that is
  $\frac{\textit{TP}}{\textit{TP+FP}}$;
\item [$\bullet$]\textit{Recall} (or also \textit{Sensitivity}): the proportion of
  real positive cases that are indeed predicted positive, that is
  $\frac{\textit{TP}}{\textit{TP+FN}}$;
\item [$\bullet$]\textit{F-Measure}: the harmonic mean of precision and recall,
  namely $\frac{2\cdot\textit{precision}\cdot
    \textit{recall}}{\textit{precision+recall}}$;
\item [$\bullet$]\textit{Matthew Correlation Coefficient} ({\it MCC} from now
  on)~\cite{Baldi2000}: the estimator of the correlation between the
  predicted class and the real class of the samples, defined as:
\[
\frac{\textit{TP}\cdot \textit{TN - FP}\cdot
      FN}{\sqrt{\textit{(TP+FN)(TP+FP)(TN+FP)(TN+FN)}}}
\]
\end{enumerate}
Each of the above measures captures a different aspect of the
  prediction quality of the samples that belong to the relevant class
  (the fake followers, in our dataset). \textit{Accuracy} measures how many samples are
  correctly identified in both of the classes, but it does not express if
  the relevant class is better recognized than the other one. Moreover,
  there are situations where some predictive models perform better
  than others, even having a lower
  accuracy~\cite{powers2011evaluation}. A high \textit{Precision} indicates
  that many of the samples identified as relevant are correctly
  recognized, but it does not give any information about the relevant samples which have not been identified. This information is provided by the
  \textit{Recall} metric, that expresses how many samples, 
  in the whole set of relevant samples,  have been correctly recognized: a low
  recall means that many relevant samples are left unidentified.
  \textit{F-Measure} and \textit{MCC} try to convey in one single
  value the quality of a prediction, combining the other
  metrics. Furthermore, \textit{MCC} is considered the unbiased version of
  the \textit{F-Measure}, since it uses all the four elements of the
  confusion matrix~\cite{powers2011evaluation}. An \textit{MCC} value
  close to 1 means that the prediction is really accurate; a value
  close to 0 means that the prediction is no better than random guess, and a
  value close to -1 means that the prediction is heavily in
  disagreement with the real class. Then, we consider as best rules
  those criteria whose application gives {\it MCC} $\ge$ 0.6, since
  such rules have the strongest correlation with the typology of the
  accounts. For completeness, when available, we also report the area-under-the-curve metric
  (\textit{AUC}), that is the area under the Receiver Operating Characteristic (ROC) curve~\cite{friedman2001elements}. The latter is the curve that
  depicts the performance of a classifier considering the percentage
  of true positive samples compared with the percentage of false
  positive samples. The \textit{AUC} is used to summarize ROC curves
  in a single value: the more the area approaches to 1, the more capable
  is the classifier.

Finally, we also report the \textit{Information Gain} (\textit{I gain}) and the
  \textit{Pearson Correlation Coefficient} (\textit{Pcc}). While the Pearson
  correlation coefficient can detect linear dependencies between a
  feature and the target class, the information gain considers a more
  general dependence, leveraging probability densities (or
  frequencies, in case of discrete variables). More precisely, the information gain is
  a measure about the informativeness of a feature with respect to the
  predicting class and it is typically adopted to train machine learning
  classifiers. It can be informally defined as the expected reduction
  in entropy caused by the knowledge of the value of a given
  attribute~\cite{Mitchell:1997:ML:541177}.
    We compute two  information gains:  \textit{I gain} about the
  outcome of the rule  and  \textit{I gain*} about the attribute
  used by the rule. For \textit{I gain}, a rule
  based on attribute $A$ can only assume the values 0 (not satisfied)
  and 1 (satisfied), while for \textit{I gain*}, the attribute $A$ 
  can assume much heterogeneous values. For example, when evaluating
  the information gain of the rule ``followers $\geq$ 30'', a sample
  with 234 followers contributes with value 1 when we compute
  \textit{I gain}, and with value 234 when we compute \textit{I
    gain$^*$}. 
The Pearson correlation coefficient, instead, is a measure of the
  strength of the linear relationship between two random variables $X$
  and $Y$~\cite{rice2006mathematical, guyon2003introduction}.
    Again , we compute
  \textit{Pcc}, considering the outcome of the satisfaction of the
  rule (namely: \textit{true}=1 or \textit{false}=0) and
  \textit{Pcc*}, based on the value assumed by the attribute used to
  evaluate the rule.  Our experiments in the following sections will
  show that, generally, a rule and the corresponding attribute assume
  very different values for the information gain and the Pearson
  correlation coefficient.

\subsection{Evaluation of Camisani-Calzolari algorithm}
\label{sec:ecc}
\begin{table}[!t]
\footnotesize
\centering
\begin{tabular}{ lrrrr }
\toprule
\multicolumn{2}{c}{} &\multicolumn{3}{c}{\textbf{outcome}}\\
\cmidrule{3-5}
\textbf{dataset}&real humans& humans & bots & neutral\\
\midrule
{\small \emph{TFP}} (@TheFakeProject) & 469 & 456 & 3 & 10 \\
{\small \emph{E13}} (\#elezioni2013)  & 1481& 1480& 0 & 1 \\
{\small \emph{FSF} $\cup$ \emph{INT} $\cup$ \emph{TWT}}\ (\textnormal{100\% fake foll.})      & 0   & 2889& 185 & 277\\
\bottomrule
\end{tabular}
\caption{Camisani-Calzolari algorithm prediction outcome.
\label{tab:cc-outcome}}
\end{table}
The detection algorithm in~\cite{Camisani:2012} aggregates the
22 criteria for identifying human and bot behavior,
introduced in Section~\ref{sec:Camisani}. The algorithm evaluates
every single rule on the account under investigation, and it assigns a
positive human score or a negative bot score, according to the output
of the rule application. The final outcome depends on the global score
obtained by the account: if the result is a score greater than 0, then
the account is marked as \textit{human}; if it is between 0 and -4, it
is marked as \textit{neutral}; otherwise, it is marked as
\textit{bot}.

Table~\ref{tab:cc-outcome} details the results of running the
algorithm over the complete dataset, including the \set{FAK}\ set,
namely all the bought fake followers accounts. Although obtaining very
good results in detecting the real human accounts, the algorithm
achieves a poor fake follower account detection. Most of the accounts
have been erroneously tagged as humans too, mainly because the fake
followers in our dataset have characteristics that easily make them
achieve a human score higher than the bot one.

The above inability to detect the fake accounts is evident in the
results of our second experiment. To evaluate the algorithm, we used
it to predict the class of the accounts of our baseline dataset
(\set{BAS}), reporting the evaluation of the final prediction in
Table~\ref{tab:ecc}. As expected, the algorithm has a poor accuracy
(very close to 0.5) and a high precision, meaning that the (few)
accounts identified as fake are effectively fake. However, it also has
a very low recall, meaning that many of the other fake accounts were
unidentified as fake. This poor performance is also expressed by a
\textit{F-Measure} close to 0.1 and by the low \textit{MCC} value.
\begin{table}[t!]
\footnotesize
\centering
\begin{tabular}{lrrrrr}
\toprule
&\multicolumn{5}{c}{\textbf{evaluation metrics}}\\
\cmidrule{2-6}
\textbf{algorithm} & \textit{accuracy} &\textit{precision}&\textit{recall}& \textit{F-M.} & \textit{MCC} \\
\midrule
CC algorithm$^{\sharp}$          &0.548	& 0.974 &0.062  & 0.116	& 0.175 \\
\bottomrule
\end{tabular}
\caption{Evaluation of Camisani-Calzolari algorithm (CC algorithm)~\cite{Camisani:2012}. Training set: \set{BAS} (1950 humans and 1950 fake followers). ($^\sharp$): the CC algorithm classified 163 accounts as neutral.
  \label{tab:ecc}}
\end{table}

\subsection{Single rule evaluation}
\label{sec:discussion}
\makeatletter{}\definecolor{Gray}{gray}{0.75}
\definecolor{LGray}{gray}{0.95}
\begin{table}[t!]
\scriptsize
\centering
\begin{tabular}{llrrrrrrrrr}
\toprule
&&\multicolumn{9}{c}{\textbf{evaluation metrics}}\\
\cmidrule{3-11}
& \textbf{rule description} & \textit{accuracy} &\hspace{-8pt}\textit{precision}&\textit{recall}& \textit{F-M.} & \textit{MCC} & \textit{I gain}& \textit{I gain$^*$}& \textit{Pcc}& \textit{Pcc$^*$}\\
\midrule
\multicolumn{8}{l}{\textit{Camisani-Calzolari}~\cite{Camisani:2012} (satisfaction of rules means human behavior)}\\
1 & profile has name                    &0.5	&--- 	&--- 	&--- 	&---           &0     &0      &0      &0      \\
2 & profile has image                   &0.496	&0.095 	&0.001 	&0.002 	&-0.06         &0.003 &0.003  &0.06   &0.06   \\
3 & profile has address                 &0.419	&0.336 	&0.166 	&0.222 	&-0.187        &0.026 &0.026  &0.187  &0.187  \\
4 & profile has biography               &0.621	&0.811 	&0.316 	&0.455 	&0.306         &0.072 &0.072  &0.306  &0.306  \\
\rowcolor{LGray}5 & followers $\ge$ 30  &0.881	&0.834 	&0.95 	&0.888 	&\textbf{0.768}&0.493 &0.525  &0.768  &0.045  \\
6 & belongs to a list                   &0.755	&0.678 	&0.971 	&0.799 	&0.566         &0.268 &0.275  &0.566  &0.127  \\
\rowcolor{LGray}7 & tweets $\ge$ 50     &0.865	&0.909 	&0.811 	&0.857 	&\textbf{0.735}&0.439 &0.621  &0.735  &0.289  \\
8 & geo-localization                    &0.667	&0.602 	&0.986 	&0.748 	&0.434         &0.163 &0.166  &0.434  &0.188  \\
9 & has URL in profile                  &0.665	&0.602 	&0.972 	&0.743 	&0.417         &0.144 &0.144  &0.417  &0.417  \\
10 & in favorites                       &0.738	&0.848 	&0.579 	&0.689 	&0.502         &0.195 &0.44   &0.502  &0.30   \\
11 & uses punctuation in tweets         &0.523	&0.979 	&0.048 	&0.091	&0.151         &0.023 &0.63   &0.161  &0.58   \\
12 & uses hashtags                      &0.608	&0.965 	&0.224 	&0.364 	&0.337         &0.1   &0.677  &0.337  &0.521  \\
13 & uses iPhone to log in              &0.705	&0.633 	&0.977 	&0.768 	&0.489         &0.202 &0.219  &0.489  &0.293  \\
14 & uses Android to log in             &0.669	&0.603 	&0.991 	&0.75 	&0.442         &0.164 &0.176  &0.446  &0.242  \\
15 & has connected with Foursquare      &0.565	&0.535 	&0.996 	&0.696 	&0.257         &0.173 &0.173  &0.442  &0.199  \\
16 & has connected with Instagram       &0.683	&0.616 	&0.969 	&0.753 	&0.446         &0.06  &0.06   &0.258  &0.144  \\
17 & uses the website Twitter.com       &0.508	&0.572 	&0.067 	&0.12 	&0.036         &0.003 &0.347  &0.061  &0.374  \\
\rowcolor{LGray}18 & has tweeted a userID&0.772	&0.992 	&0.549 	&0.707 	&\textbf{0.609}&0.334 &0.752  &0.609  &0.544  \\
19 & 2*followers $\ge$ friends        	&0.721	&0.642 	&0.998 	&0.781 	&0.531         &0.26  &0.26   &0.531  &0.531  \\
20 & tweets do not just contain URLs    &0.53	&0.947 	&0.064 	&0.12 	&0.167         &0.299 &0.673  &0.587  &0.403  \\
21 & retwitted tweets $\ge$ 1           &0.753	&0.967 	&0.524 	&0.679 	&0.569         &0.278 &0.722  &0.569  &0.49   \\
22 & uses different clients  to log in  &0.524	&0.819 	&0.061 	&0.113 	&0.125         &0.018 &0.629  &0.144  &0.55   \\
\midrule

\multicolumn{8}{l}{\textit{Van Den Beld (State of search)}~\cite{blogger1:2012} (satisfaction of rules means bot behavior)}\\
1 & {\it bot} in biography              &0.5	& ---   & ---	  & ---	 & --- & 0      &0    & 0      &0      \\
2 & following:followers = 100:1         &0.541	& 0.541 & 1 	  & 0.15 &0.205& 0      &0    & 0.006  &0.006  \\
3 & same sentence to many accounts      &0.438	& 0.438 & 0.78  & 0.146& -0.169& 0.444  &0.444& 0.74   &0.74   \\
4 & duplicate profile pictures          &0.471	& 0.471 & 0.928 & 0.025& -0.146& 0      &0    & 0      &0      \\
5 & tweet from API                      &0.118	& 0.118 & 0.017 & 0.2  & -0.779& 0.528  &0.694& 0.779  &0.505  \\
\midrule                         

\multicolumn{8}{l}{\textit{Socialbakers}~\cite{SocialBakers} (satisfaction of rules means fake behavior)}\\
1 & friends:followers $\ge$ 50:1           & 0.581  &0.997 &0.162 &0.279 &0.296  &0      &0      &0.011  &0.011\\  
2 & tweets spam phrases                    & 0.501  &1     &0.003 &0.005 &0.036  &0.435  &0.45   &0.719  &0.404\\
3 & same tweet $\ge$ 3                     & 0.348  &0.046 &0.015 &0.023 &-0.407 &0.166  &0.317  &0.441  &0.284\\  
4 & retweets $\ge$ 90\%                    & 0.499  &0.452 &0.007 &0.014 &-0.009 &0      &0.478  &0.007  &0.555\\  
5 & tweet-links$\ge$ 90\%                  & 0.511  &0.806 &0.03  &0.057 &0.084  &0.006  &0.401  &0.087  &0.353\\
6 & 0 tweets                               & 0.521  &0.988 &0.043 &0.083 &0.146  &0.02   &0.621  &0.146  &0.289\\  
7 & default image after 2 months           & 0.496  &0.095 &0.001 &0.002 &-0.06  &0.003  &0.003  &0.06   &0.06 \\
8 & no bio, no location, friends $\ge$100  & 0.559  &0.917 &0.131 &0.229 &0.231  &0.004  &0.004  &0.079  &0.079\\
\bottomrule
\end{tabular}
\caption{Rules evaluation.}
\label{table:results}
\end{table}

  In this section, we analyze the
  effectiveness of each single rule, as designed by the original
  authors, in order to evaluate which rule can be considered as a good
  criterion for the detection of fake Twitter followers.

Table~\ref{table:results} summarizes the results obtained by the
application of each rule introduced in Sections~\ref{sec:Camisani},
\ref{app:blogs}, and \ref{app:online-analyzers} to our \set{BAS}\
dataset. In Table~\ref{table:results}, we highlighted the rules achieving high {\it MCC} values. As shown, only three
rules obtained a value higher than 0.6, namely: (1) the threshold of
at least 30 followers, (2) the threshold of at least 50 tweets, and
(3) the use of a userID in at least one tweet.

As expected by the definition of \textit{MCC}, such rules also exhibit
a combination of high accuracy, precision, and recall.
However, it is worth observing the values for the information gain
  and the Pearson correlation coefficient. The information gain of the
  rules (\textit{I gain}) is always lower than the evaluation of the
  related attribute \textit{I gain*}, while this is not true for the
  Pearson correlation coefficient (\textit{Pcc} and
  \textit{Pcc*}). Actually, this happens because \textit{Pcc}
  evaluates the linear dependency between two variables that assume
  very similar values, namely the output of the rule and the class,
  while the \textit{Pcc*} considers variables with more heterogeneous
  values. In the first case, indeed, both the variables class and the
  output can assume only the values 0 and 1: the class can be either 0
  (\textit{human}) or 1 (\textit{fake}), the rules can output either 0
  (\textit{false}, for example, \textit{account does not have more
    than 50 tweets}) or 1 (\textit{true}, for example, \textit{account
    has more than 50 tweets}). Instead, for the \textit{Pcc*} , the
  attribute of a rule (in the example, the number of tweets) can
  assume much higher values (\textit{account has 234 tweets}). This is
  clearly not linearly dependent on the class values,
  resulting in lower
  values for the \textit{Pcc*}  with respect to the
  \textit{Pcc}~\cite{rice2006mathematical}. 

Thus, for each rule listed in Section~\ref{sec:Camisani} (top part
  of Table~\ref{table:results}), it is meaningless to compare the
  \textit{Pcc} and \textit{Pcc$^*$} values. Instead, we need to focus
  only on the same type of metric, namely by column, to compare the
  linear dependency of the feature with the class.
  Then, directing our attention to the information gain, we notice
  that many of the rules take into account attributes that are
  effectively able to perform the discrimination between the two
  classes. If we consider as useful the rules and features that have
  an information gain value higher than 0.5, we observe that, even if
  many rules exhibit a very low \textit{\mbox{I gain}}, their
  ``feature" version becomes much more interesting: for example, rules
  18, 20, 21 and 22 have an evident increase in their information gain
  when used as features. Thus, we can derive that the rule is based on
  a right assumption (for example, the use of hashtags), but the rule
  definition is too simple to be effective: the algorithm proposed
  by~\cite{Camisani:2012} is simply too naive for the complex task of
  fake accounts detection in Twitter. Coincidentally, we have that the
  best performing rules also show the highest \textit{Pcc} values,
  namely their satisfaction is more strongly related to the belonging
  class. Concerning the features underlying the rules, we find that
  the \textit{Pcc$^*$} is strongly reduced because, as above noticed,
  they can (and indeed do) assume very high values and this severely
  affects the linear correlation with the class.

Observing the other rules of Table~\ref{table:results}, we can
notice how none of the criteria suggested by online blogs and
by Socialbakers' FakeFollowerCheck are successful in detecting the fake
followers in our dataset. 
As an example, all rules by Van Den Beld have accuracy and precision
close to 0.5 or a very low recall.  Also, we observe that ``tweet from
API'' has an \textit{MCC} of -0.779, meaning that it is strictly
related to the class of the account, but by an inverse factor: in our
dataset, fake followers accounts almost never tweet from API (instead,
they use Twitter.com to tweet), whereas human accounts have posted at
least once from outside the website. This is exactly the opposite
behavior than that suggested by the blogger for bots, that (are
supposed to) almost exclusively post tweets using API. The
  relevance to the prediction task is also confirmed by both the
  \textit{I gain/I gain$^*$} and the \textit{Pcc/Pcc$^*$} values.

Another interesting observation is that many rules proposed by
Socialbakers have \textit{MCC} values close to 0, meaning that their
outcomes are almost unrelated with the class of the accounts. Indeed,
the large majority of the accounts are recognized as humans, resulting
in a high precision, accuracy around 0.5 and very low recall. The
  exception is rule 6, ``0 tweets'': as a rule, it has an
  information gain value of 0.02, but when considered as a feature
  (i.e., number of tweets) it obtains 0.621. Similarly, rules 4 and 5 are
  much more useful for the detection process when considering their
  underlying features (namely, number of retweets and number of
  tweets with URLs).
Summarizing, independently from the typology of the account, the rules
are almost always satisfied, leading to a severe flaw when dealing
with fake followers detection.

\makeatletter{}\section{Fake detection with  algorithms based on feature
  sets} \label{sec:spamdet}
In this section, we examine works in~\cite{Stringhini:2010,Yang:2013}
that address spam account detection on Twitter.  Both of them propose
a list of features to be extracted from manually classified datasets
of accounts. Such feature sets are then used to train and test machine
learning classifiers in order to distinguish between humans and
spammers. Even if the proposed features have been originally designed
for spam detection, here, for the first time, we consider them to spot
another category of Twitter accounts, i.e., the fake
followers. Although many other works exist in literature focused on
Twitter spam detection (see Section~\ref{sec:RW}), many of them
consider features that can be in some way assimilated to those
analyzed in this and in the previous section.

 Differently from the rule-based algorithms in Section~\ref{sec:fakedet}, 
  features are here presented as quantifications of properties of the
  considered samples. Therefore, they are introduced without any prior
  knowledge about the values for the features that will characterize the
  considered classes. Only after the training phase, it will be possible to
  observe which are the most frequent values for the features within
  the different classes.

 For our analysis, we employ classifiers that produce both ``glass-box'' and ``black-box'' models.
In ``glass-box'' models, such as Decision Trees and Regression Models, the inner structure of the models can be understood by humans, also providing insights on how the classifiers identify fake accounts~\cite{friedman2001elements}.
Instead, in ``black-box" models, such as Support Vector Machines, the inner structure of the model does not have a direct human-explicable correspondence.

\subsection{Detecting spammers in social
  networks}\label{detailsString10}
The study presented in~\cite{Stringhini:2010} focuses on spambot
detection. The authors exploit several characteristics that can be gathered
crawling an account's details, both from its profile and timeline. For each investigated account, such characteristics are exploited in a
Random Forest algorithm~\cite{data_mining,weka}, that outputs if the
account is a spambot or not.
\label{app:stringhini}
The results of the analysis in~\cite{Stringhini:2010} depicted
  some interesting features of the spambot accounts under
  investigation, as reported in Table~\ref{Stringhinifeatset}.

\begin{table}[h!]
  \footnotesize
    \begin{tabular}{p{0.45\textwidth}p{0.55\textwidth}}
	\toprule
	\textbf{1.}   spambots do not have thousands of friends\label{item:2};	& \textbf{4.}  spambots have a high $\frac{\textit{tweets containing URLs}}{\textit{total tweets}}$ ratio (\textit{URL ratio})\label{item:5}; \\
	\textbf{2.}   spambots have sent less than 20 tweets\label{item:3}; & \textbf{5.}  spambots have a high $\frac{\textit{friends}}{(\textit{followers\string^2})}$ ratio value (i.e., lower ratio values mean legitimate users)\label{item:6}. \\
	\textbf{3.}   the content of spambots' tweets exhibits the so-called {\it message similarity}\label{item:4};		& \\
	\bottomrule
    \end{tabular}
\caption{Feature set proposed by Stringhini \textit{et al.}~\cite{Stringhini:2010}.
\label{Stringhinifeatset}}
\end{table}

To evaluate feature 3, we implement the notion of {\it message similarity} by checking the existence of at least two tweets, in the
last 15 tweets of the account timeline, in which 4 consecutive words
are equal. This notion has been given in a later work by the same
authors~\cite{Stringhini:2012}.

Without the original training set, we were unable to reproduce the
same classifier, but we picked the five features and used them to
train a set of classifiers with our \set{BAS}\ dataset. The results
are reported in Table~\ref{table:classifiersSezione5} of
Section~\ref{subsec:evaluationspammers}.

\subsection{Fighting evolving Twitter
  spammers\cut{~\cite{Yang:2013}}}\label{app:yang}
The authors of~\cite{Yang:2013} observed that Twitter
spammers often modify their behavior in order to evade existing spam
detection techniques. Thus, they suggested to consider some new
features, making evasion more difficult for spammers. Beyond the features directly available from the account profile
lookup, the authors propose some \mbox{graph-,} automation\mbox{-,}
and timing-based features. In Table~\ref{Yangfeatset} we detail nine of them, together with the outcome of their analysis in~\cite{Yang:2013}.

\begin{table}[h!]
  \footnotesize
    \begin{tabular}{p{0.50\textwidth}p{0.50\textwidth}}
	\toprule
	\textbf{1.}   {\it age} of the account (this feature also appears in~\cite{Lee:2010}): the more an account is aged, the more it could be considered a good one;	& \textbf{6.}  {\it API ratio} ($\frac{\textit{tweets sent from API}}{\textit{total number of tweets}}$): higher values for suspicious accounts; \\
	\textbf{2.}   {\it bidirectional link ratio} ($\frac{\textit{bidirectional links}}{\textit{friends}}$), where a bidirectional link occurs when two accounts follow each other: this feature has been tested to be lower for spammer accounts than for legitimate accounts; & \textbf{7.}  {\it API URL ratio} ($\frac{\textit{tweets sent from API and containing URLs}}{\textit{total number of tweets sent from API}}$): such ratio is  higher for suspicious accounts; \\
	\textbf{3.}   {\it average neighbors' followers}: the average number of followers of the account's friends. This feature aims at reflecting the quality of the choice of friends of an account. The feature has been found to be commonly higher for legitimate accounts than for spammers;		& \textbf{8.}  {\it API tweet similarity}: this metric considers only the number of similar tweets sent from API. The notion of tweet similarity is as in Section~\ref{app:stringhini}. This metric is higher for suspicious accounts;\\
	\textbf{4.}   {\it average neighbors' tweets}: the average number of tweets of the account's followers. This feature is lower for spammers than for legitimate accounts;		& \textbf{9.}  {\it following rate}: this metric reflects the speed at which an accounts follows other accounts\label{item:1}. Spammers usually feature high values of this rate. \\
	\textbf{5.}   {\it followings to median neighbor's followers}: defined as the ratio between the number of friends and the median of the followers of its friends. This feature has been found higher for spammers than for legitimate accounts;		& \\
	\bottomrule
    \end{tabular}
\caption{Feature set proposed by Yang \textit{et al.}~\cite{Yang:2013}.
\label{Yangfeatset}}
\end{table}

The authors of~\cite{Yang:2013} combine their features in
four different machine learning classifiers and compare their
implementation with other existing approaches.  We were unable to completely reproduce the machine learning
classifiers in~\cite{Yang:2013}, since we had a different
dataset. Instead, here we evaluate how those features, which proved to
be quite robust against evasion techniques adopted by spammers,
perform in detecting fake Twitter followers.
As in~\cite{Yang:2013}, the following rate (feature 9) has
been approximated with the ratio \textit{friends/age}, since a precise
evaluation would require to know the evolution of the number of
friends of an account, but this is, indeed, publicly unavailable.

Finally, in~\cite{Yang:2013} there are also other features
in addition to those mentioned above. However, as claimed by the same
authors, they are less robust with regards to evasion techniques and thus we decided not to include them in our evaluation.

\subsection{Evaluation }\label{subsec:evaluationspammers}
\makeatletter{}\begin{table}[t!]
  \footnotesize
  \centering
  \begin{tabular}{llrr}
    \toprule
    &&\multicolumn{2}{c}{\textbf{evaluation metrics}}\\
    \cmidrule{3-4}
        & \textbf{feature description} & \textit{I gain}& \textit{Pcc}\\
    \midrule
\multicolumn{4}{l}{\textit{Stringhini et al.}~\cite{Stringhini:2010}}\\
1  &number of friends     &0.263  &0.030 \\
2  &number of tweets      &0.621  &0.289\\
3  &content of tweets     &0.444  &0.740 \\
4  &URL ratio in tweets   &0.401  &0.353\\
5  &$\frac{\textit{friends}}{(\textit{followers\string^2})}$&0.733  &0.169\\
\midrule
\multicolumn{4}{l}{\textit{Yang et al.}~\cite{Yang:2013}}\\
1  &age                        &0.539  &0.436       \\
2  &bidirectional links ratio  &0.905  &0.875       \\
3  &avg. followers of friends  &0.327  &0.254       \\
4  &avg. tweets of friends     &0.203  &0.235       \\
5  &firends / med. foll. of friends  &0.336  &0.102 \\
6  &api ratio                  &0.544  &0.635       \\
7  &api url ratio              &0.058  &0.113       \\
8  &api tweet similarity       &0.460  &0.748        \\
9  &following rate             &0.355  &0.214       \\
\bottomrule
\end{tabular}
\caption{Evaluation of the single feature.}
\label{table:feature:set}
\end{table}

As done for the rule set in Section~\ref{sec:fakedet}, we report
  in Table~\ref{table:feature:set} the evaluation of the information
  gain and the Pearson correlation coefficient for all the features
  within the \emph{BAS} dataset. Also in this case, since the
  \textit{Pcc} evaluates the linear dependence between the considered
  feature and the class (that can only be 0 or 1), it produces results
  that are slightly different when compared to the information
  gain. Observing the results in Table~\ref{table:feature:set}, we can
  identify several promising features: ``number of tweets'' (already
  noticed in Section~\ref{sec:fakedet}), ``ratio between friends and
  followers$\string^2$'', ``bidirectional links ratio'' and ``API
  ratio''. The beneficial effect of the bi-link ratio will be further
  confirmed by the experiments of Section~\ref{sec:bilink-ratio}.

To evaluate the combined effectiveness of the feature sets described in Sections~\ref{detailsString10} and~\ref{app:yang} on detecting fake follower accounts, we employed 8 classifiers, obtained from different machine learning-based algorithms, namely: Decorate (D), Adaptive Boost (AB), Random Forest (RF), Decision Tree (J48), Bayesian Network (BN), k-Nearest Neighbors (kNN), Multinomial Ridge Logistic Regression (LR) and a Support Vector Machine (SVM). Our SVM classifier exploits a Radial Basis Function (RBF) kernel and has been trained using libSVM as the machine learning algorithm~\cite{chang2011libsvm}. During the training phase of the SVM, the \textit{cost} and \textit{gamma} parameters have been optimized via a grid search algorithm. Similarly, the \textit{k} parameter of the kNN classifier and the \textit{ridge} penalizing parameter of the LR model have been optimized via a cross validation parameter selection algorithm. All the classifiers and the optimization algorithms employed in this work are implemented within the Weka framework~\cite{weka}.

Among these algorithms, RF was the only one used in~\cite{Stringhini:2010}. Instead, authors of~\cite{Yang:2013} employed D, RF, J48 and BN. We have decided to include AB in our work, since it is considered one of the most effective machine learning algorithms for classification tasks~\cite{friedman2001elements}. Furthermore, we also added other well-known and widely adopted classifiers, which are based on different classification techniques, such as SVM, kNN and LR, in order to perform a thorough evaluation of our detection system.We have built 8 classifiers adopting the features in Sections~\ref{detailsString10} and~\ref{app:yang} and we have trained the models using our baseline (\textit{BAS}) dataset. Then, we have used a 10-fold cross validation~\cite{data_mining} to estimate the performances of each obtained classifier.  As for the rule-based algorithms in Section~\ref{sec:evaluation}, we look at the \textit{MCC} as the preferred metric to assess the classifiers' performances. Table~\ref{table:classifiersSezione5} summarises the results. The highest values for each metric are shown in bold.

\label{sec:class-perf}
\makeatletter{}\begin{table}[t!]
\footnotesize
\centering
\begin{tabular}{llrrrrrr}
\toprule
\cellcolor{white}&&\multicolumn{6}{c}{\textbf{evaluation metrics}}\\
\cmidrule{3-8}
&\textbf{algorithm} & \textit{accuracy} &\textit{precision}&\textit{recall}& \textit{F-M.} & \textit{MCC} & \textit{AUC}\\
\midrule
\multicolumn{6}{l}{\textit{Classifiers based on feature set by Yang et al. \cite{Yang:2013}}}\\
RF&Random Forest 	  &\textbf{0.991}  &0.991  &0.991  &\textbf{0.991}  &\textbf{0.983}  &0.998\\
D&Decorate            	  &\textbf{0.991}  &0.988  &\textbf{0.993}  &\textbf{0.991}  &\textbf{0.983}  &0.998\\
J48&Decision Tree	  &0.990  &0.991  &0.989  &0.990  &0.980  &0.997\\
AB&Adaptive Boosting	  &0.988  &0.989  &0.987  &0.988  &0.974  &\textbf{0.999}\\
BN&Bayesian Network	  &0.976  &\textbf{0.994}  &0.958  &0.976  &0.936  &0.997\\
kNN&k-Nearest Neighbors	  &0.966  &0.966  &0.966  &0.966  &0.932  &0.983\\
LR&Logistic Regression	  &0.969  &0.966  &0.973  &0.969  &0.939  &0.996\\
SVM&Support Vector Machine	  &0.989  &0.985  &\textbf{0.993}  &0.989  &0.976  &0.989\\
\midrule
\multicolumn{6}{l}{\textit{Classifiers based on feature set by Stringhini et al. \cite{Stringhini:2010}}}\\
RF&Random Forest 	&  0.981  &0.983  &\textbf{0.979}  &\textbf{0.981}  &\textbf{0.961}  &\textbf{0.995}\\
D&Decorate            	&  \textbf{0.982}  &\textbf{0.984}  &\textbf{0.979}  &\textbf{0.981}  &\textbf{0.961}  &0.993\\
J48&Decision Tree	&  0.979  &\textbf{0.984}  &0.974  &0.979  &0.953  &0.985\\
AB&Adaptive Boosting	&  0.968  &0.965  &0.970  &0.968  &0.938  &\textbf{0.995}\\
BN&Bayesian Network	&  0.953  &0.953  &0.953  &0.953  &0.907  &0.985\\
kNN&k-Nearest Neighbors	  &0.954  &0.961  &0.946  &0.953  &0.907  &0.974\\
LR&Logistic Regression	  &0.927  &0.921  &0.935  &0.928  &0.855  &0.974\\
SVM&Support Vector Machine	  &0.959  &0.967  &0.950  &0.958  &0.917  &0.958\\
\bottomrule
\end{tabular}
\caption{Performance comparison for 10-fold cross validation. Training set: \set{BAS}.}
\label{table:classifiersSezione5}
\end{table}

We can observe that all the classifiers have an excellent prediction capability. The ones built over the feature set of~\cite{Yang:2013} achieve slightly better results.
In particular, the RF, J48 and D classifiers have \textit{MCC}  greater than 0.98. Similarly, precision and recall are around 0.99 for all of them.
In addition, all the classifiers based on the feature set by~\cite{Yang:2013} have a higher \textit{AUC}, when compared with the ones built with the feature set by~\cite{Stringhini:2010}.
Anyway, the latter also obtain high detection levels: accuracy, precision, and recall are around 0.98 for RF, D and J48, with a \textit{MCC} of around 0.96. The lower precision and recall with respect to the ones obtained using the set of Yang \textit{et al.}~\cite{Yang:2013} show that the features of Stringhini \textit{et al.}~\cite{Stringhini:2010} exhibit the tendency to consider as fake followers some human accounts.
With both the ~\cite{Yang:2013} and~\cite{Stringhini:2010} feature sets, BN, kNN and LR classifiers achieve, overall, worse performances. The SVM classifier, instead, achieves remarkable results, especially with the feature set of~\cite{Yang:2013}. Indeed, in this experiment SVM scores only slightly worse than RF, D and J48, and better than AB. Whereas, AB achieves extremely high performances when evaluated with the \textit{AUC} metric. Finally, among all the considered classifiers and evaluation metrics, RF and D are the ones that have been proved to be more consistent.

Overall, even if some small differences can be observed in the evaluation metrics, all the classifiers almost correctly distinguish between human and fake follower accounts, for our baseline \set{BAS}\ dataset.
The feature-based classifiers are indisputably more accurate for fake follower detection when compared with the CC algorithm, that does not perform well within our dataset, as observed in Section~\ref{sec:ecc} above.

\subsection{Discussion}\label{sec:discussion-1}

By examining the internal structure of the classifiers, we get
insights about the best features that contribute more to distinguish between 
human and fake followers. In the case of decision trees, the best features are the
ones closer to the root and the classifier automatically finds the
numeric thresholds characterizing, for a given feature, the borderline
between human and fake followers. It is worth noting that also the Decorate,
AdaBoost, and Random Forest algorithms exploit, ultimately,
combinations of simple decision tree classifiers.  Despite their very
good performance, they have the disadvantage of being difficult to
analyze, since they can consist in tens of individual trees that
interact together. Then, we only focus on the J48 classifier (a single
decision tree) to examine how the features are applied during the
classification process.

\subsubsection{Differences between fake followers and spam accounts}
\label{sec:diff-betw-fake}
Looking at the tree structure, we observe some interesting
differences between the fake followers in our \set{BAS}\ dataset and the spam accounts
characterized in~\cite{Stringhini:2010} and~\cite{Yang:2013}. For example,
the feature \textit{URL ratio} has been found to have a higher value for spammers than for legitimate users, as highlighted in~\cite{Stringhini:2010} 
(Section~\ref{app:stringhini}). Observing the tree structure of our
J48 classifier, instead, low values for this feature characterize fake followers, compared with higher values that indicate human accounts in our baseline dataset. 
More than 72\% of the fake followers in our training dataset have a
\textit{URL ratio} lower than 0.05, oppositely to 14\% of human
accounts. Similarly, the \textit{API ratio} feature has been found higher for spammers than for legitimate accounts (\cite{Yang:2013}, see also Section~\ref{app:yang}). In our dataset,  the \textit{API ratio} is lower than 0.0001 for 78\%
of fake followers. A similar behavior has been observed for the
\textit{average neighbor's tweets} feature, that has been found to be lower  for spammers in~\cite{Yang:2013},
but higher for our fake followers.

These initial observations highlight a behavioral difference between
a spam account and a fake follower. In particular, fake
followers appear to be more passive compared to  spammers and they do 
not make use of automated mechanisms for posting their tweets, as
 spammers usually do.

\subsubsection{Reducing overfitting}\label{sec:reducing-overfitting}
\makeatletter{}\begin{table*}[t!]
\footnotesize
  \centering
  \begin{tabular}{lrrr@{\phantom{M}}rrrrrrr}
    \toprule
    &\multicolumn{3}{c}{\textbf{tree details}}
      &&\multicolumn{6}{c}{\textbf{evaluation metrics}}\\
    \cmidrule{2-4}\cmidrule{6-11}
\textbf{pruning method} & \textit{nodes} & \textit{leaves} & \textit{height} & & \textit{accuracy}& \textit{precision}&\textit{recall}& \textit{F-M.}&\textit{MCC}&\textit{AUC}\\
\midrule
\multicolumn{8}{l}{\textit{Decision tree based on feature set of Stringhini et al.\cite{Stringhini:2010}}}\\
subtree raising 0.25 &   43&     22&     7&&0.979  &0.984  &0.974  &0.979  &0.953  &0.985\\
reduced error 3 folds&   31&     16&     5&&0.975  &0.971  &0.971  &0.971  &0.943  &0.989\\
reduced error 50 folds&   9&      5&     4&&0.964  &0.957  &0.957  &0.957  &0.914  &0.984\\
\midrule
\multicolumn{9}{l}{\textit{Decision tree based on feature set of Yang et al.\cite{Yang:2013}}}\\
subtree raising 0.25 & 33&     17&     8&& 0.99  &0.991  &0.989  &0.99  &0.98  &0.997  \\
reduced error 3 folds& 19&     10&     5&& 0.988  &0.99  &0.987  &0.988  &0.976  &0.993  \\
reduced error 50 folds&11&      6&     3&& 0.982  &0.979  &0.985  &0.982  &0.966  &0.993  \\
\midrule
\multicolumn{9}{l}{\textit{Decision tree based on feature set of Yang et al.\cite{Yang:2013}, without the bi-link ratio feature}}\\
subtree raising 0.25 &  101&     51&   10&&0.96   &0.963  &0.957  &0.96   &0.917  &0.971\\
reduced error 3 folds&   53&     27&    8&&0.961  &0.969  &0.952  &0.96   &0.914  &0.978\\
reduced error 50 folds&  37&     19&    9&&0.933  &0.931  &0.934  &0.933  &0.866  &0.967\\
\bottomrule
\end{tabular}
\caption{Performance comparison with increased pruning. 10-fold cross validation. Training set: \set{BAS}.}
\label{table:generalize}
\end{table*}

It is well known that trained classifiers can be subject to
``overfitting'', namely the problem of being too specialized on the
training dataset and unable to generalize the classification to new
and unseen data~\cite{hawkins2004problem}.

A simple way to avoid overfitting is to keep the classifier as simple
as possible. In case of a decision tree algorithm, for example, one
solution could be reducing the number of nodes and, possibly, the
height of the tree.  The decision tree obtained with the feature set
of Stringhini \textit{et al.}~\cite{Stringhini:2010} has 22 leaves, 43 nodes,
and a height of 7, whereas the best feature is the
\textit{friends/(followers\string^2)} ratio that places at the
root. The decision tree with the feature set of Yang \textit{et
al.}~\cite{Yang:2013} has 17 leaves, 33 nodes and a height of 8, with
the \textit{bi-directional link} ratio as the root.

A common practice to generalize the classifiers is the adoption of a
more aggressive pruning strategy, e.g., by using the reduce-error
pruning with small test sets~\cite{data_mining,weka}. Adopting this
strategy, we were able to obtain simpler trees with a lower number of
nodes and a very reduced height. Such simpler trees generally use subsets
of the feature set, still maintaining very good performance on our
\set{BAS}\ dataset.

Table~\ref{table:generalize} reports the characteristics and the
performance of the experiments we have carried out, varying the
pruning strategy. It is worth noting that the complexity of the tree
is not always directly connected to an improvement in the detection
capability: for example, for the feature set of Yang \textit{et al.}~\cite{Yang:2013}, reducing the number of nodes from 33 to 11
decreases the accuracy of 0.007 and the \textit{MCC} of 0.014,
only. Similarly, the values for \textit{AUC} remain almost the
  same between the pruned and the not pruned versions of the tree.
    Furthermore, we clearly observe that the
  pruned version of Stringhini \textit{et al.}~\cite{Stringhini:2010} reduces its recall of 0.017,
  while the one of Yang \textit{et al.}\cite{Yang:2013} only drops of 0.004, meaning that the
  latter is able to miss fewer fakes than the former one after
  pruning. This is also evident from the higher reduction of both
  \textit{F-Measure} and \textit{MCC}. We think that this increased
  effectiveness is a direct consequence of the quality  of
  the used features.  Overall, the results of this experiment show
that, even reducing the features, it is possible to have a detection
rate higher than 0.95 (as in the last lines of
Table~\ref{table:generalize}, for~\cite{Stringhini:2010}
and~\cite{Yang:2013}, respectively). For instance, in those two
experiments, the features used by the pruned tree were only
\textit{bi-directional link ratio}, the \textit{average neighbors'
  followers}, the \textit{age}, and the \textit{followings to median
  neighbors' followers} as a subset of the original feature set of
Yang \textit{et al.}~\cite{Yang:2013}, and the
\textit{friends/(followers\string^2)}, \textit{URL ratio}, and
\textit{number of friends} as the subset for the Stringhini \textit{et al.}~\cite{Stringhini:2010} original feature set.

\subsubsection{\textit{Bidirectional link ratio}}\label{sec:bilink-ratio}
In Section~\ref{subsec:evaluationspammers} we observed that the
\textit{bidirectional link ratio} had the highest information gain
among all the considered features. In order to test if this is the
decisive feature to distinguish between humans and fake followers and how much it influences the detection
process, we compare the results of the previous experiments with those
of a new one. We build a decision tree classifier leaving out the
bi-link ratio from the feature set of Yang \textit{et
  al.}~\cite{Yang:2013} and compare its effectiveness against the
classifier built with the complete set. The results are reported in
the last rows of Table~\ref{table:generalize}.

This experiment is particularly interesting since, as detailed in next
Section~\ref{sec:opt-cost}, this feature is the most expensive to
evaluate.
The results in Table~\ref{table:generalize} show an evident decrease
in \textit{accuracy}, \textit{precision} and \textit{recall} for the
less pruned trees (subtree raising 0.25 and reduced error 3 folds), as
well as for both \textit{F-Measure} and \textit{AUC}, and an even
noticeable decrease of the \textit{MCC} measure. The reduced error
pruning with 50 folds produces a classifier that has \textit{MCC}
dropping from 0.966 to 0.866. Its detection level (\textit{accuracy})
is still very good for all the three pruned trees (0.964, 0.982 and
0.933, respectively), but we can clearly observe a remarkable drop in
both precision and recall, compared to the version with the whole
feature set. This suggests that the highest effectiveness we noticed
in the above experiments after pruning
(Section~\ref{sec:reducing-overfitting}) is considerably lost when we
do not consider the bi-link ratio feature.  The most interesting
aspect is the increased complexity of the decision tree: without the
bi-link ratio, the classifiers need to resort to a considerably larger
number of nodes. For example, the tree that does not use that feature,
pruned with subtree raising confidence of 0.25, requires 101 nodes,
whereas the tree that uses it requires only 33 nodes.

From the results shown in Table~\ref{table:generalize}, we conclude
that the bidirectional link ratio is an important feature for fake
follower detection: even if not essential, it is extremely effective
for the detection process. By capturing the nature of the social
  ties between an account and its neighbors, this feature is
  intrinsically harder to beat than those based on simpler
  characteristics, like, e.g., other information in the account's
  profile.

\makeatletter{}\section{An efficient and lightweight classifier}\label{sec:opt-cost}
\makeatletter{}\newcolumntype{C}[1]{>{\centering\arraybackslash}m{#1}}
\begin{table*}[ht!]
\scriptsize
  \centering
  \begin{tabular}{lC{0.27\textwidth}C{0.25\textwidth}C{0.22\textwidth}}
    \toprule
    \textbf{Feature set}&  \textbf{Class A (profile)}&    \textbf{Class B (timeline)}&    \textbf{Class C (relationships)}\\
    \midrule
    \textit{\footnotesize Camisani-Calzolari}~\cite{Camisani:2012}&
    has name, has image, has address, has biography, followers $\ge$ 30, belongs to a list, tweets $\ge$ 50, URL in profile, 2 $\times$ followers $\ge$ friends
    & geo-localized, is favorite, uses punctuation, uses hashtag, uses iPhone, uses Android, uses Foursquare, uses Instagram, uses \textit{Twitter.com}, userID in tweet, tweets with URLs, retweet $\ge$ 1, uses different clients&\\
    \midrule
    \textit{\footnotesize State of search}~\cite{blogger1:2012}&
    {\it bot} in biography, $\frac{\textit{friends}}{\textit{followers}}$ $\simeq$ 100, duplicate profile pictures 
    & same sentence to many accounts, tweet from API &\\
    \midrule
    \textit{\footnotesize Socialbakers}~\cite{SocialBakers} &
    $\frac{\textit{friends}}{\textit{followers}}$ $\ge$ 50, default image after 2 months, no bio, no location, friends $\ge$100, 0 tweets 
    & tweets spam phrases, same tweet $\ge$ 3, retweets $\ge$ 90\%, tweet-links $\ge$ 90\% &\\
    \midrule
    \textit{\footnotesize Stringhini et al.}~\cite{Stringhini:2010} &
    number of friends, number of tweets, $\frac{\textit{friends}}{\textit{(followers\string^2)}}$ 
    & tweet similarity, URL ratio &\\
    \midrule
    \textit{\footnotesize Yang et al.}~\cite{Yang:2013}& age, following rate
    & API ratio,  API URL ratio,  API tweet similarity
    & bi-link ratio, average neighbors' followers, average neighbors' tweets, followings to median neighbor's followers\\
    \bottomrule
\end{tabular}
\caption{Feature crawling cost classes.}
\label{tab:feature-categories}
\end{table*}

As previously shown in sections~\ref{sec:fakedet}
and~\ref{sec:spamdet}, the classifiers based on feature sets perform
much better than those based on rules. Similarly, we have seen that
the feature set proposed by Yang \textit{et al.}~\cite{Yang:2013}
seems to be slightly more effective than the one proposed by Stringhini
\textit{et al.}\cite{Stringhini:2010}, when used in feature-based
classifiers aiming at fake followers detection.
Here, we look for an efficient and lightweight classifier, exploiting the best
features and the best rules, not only in terms of detection
performance, but also considering their evaluation costs. In
particular, we can distinguish between the computational cost and the
crawling cost required to evaluate a feature (or a
rule). Computational costs can be generally lowered with optimized
algorithms and data representations and they are negligible when
compared to the crawling costs. Thus, in this section we focus on the
latter: we quantify the crawling cost of each feature and rule, and we
build a set of lightweight classifiers that make use of the most
efficient features and rules, in terms of crawling cost and fake
followers detection capability. For the sake of readability, in the following section with the term
``feature'' we  include all the features presented in
sections~\ref{sec:fakedet} and~\ref{sec:spamdet}, namely also the features
underlying the analyzed rules.

\subsection{Crawling cost analysis}
\label{sec:costs}
\newcommand{\foll}{\ensuremath{\textit{f}}}
\newcommand{\fri}{\ensuremath{\varphi}}
Intuitively, some features require few data for their calculation,
while others require the download of big amounts of data.  For the sake of this analysis, we divide the features in three
categories:
\begin{enumerate}
\item [\textit{A})]\textit{profile}: features that require information
  present in the profile of the followers of the target account (like, e.g.,
  \textit{profile has name});
\item [\textit{B})]\textit{timeline}: features that require the tweets
  posted in the timeline of the followers of the target account (like, e.g.,
  \textit{tweet from API});
\item [\textit{C})]\textit{relationship}: features that require
  information about the accounts that are in a relationship (i.e., that are a 
  friend, or a follower, or both) with the followers of the target account 
   (like, e.g., \textit{bidirectional link ratio}).
 \end{enumerate}
 Each category, in turn, belongs to a crawling cost class directly
related to the amount of data to be crawled from
Twitter. Starting from the list of the followers of a target account,
\textit{Class A} features can be evaluated simply accessing to all the
profiles of the followers; \textit{Class B} features require to
download all the tweets posted by each follower; \textit{Class C}
features need to crawl the friends and the followers of each follower
of the target account. To evaluate the class of cost associated to each feature's category,
we estimate the number of Twitter \textit{API calls} needed to download data
required for the calculation.  Results are in
Tables~\ref{tab:singleAbs} and~\ref{tab:feature-categories}.
The following parameters refer to the Twitter account for which the
number of fake followers is being  investigated:
\begin{enumerate}
\item [\foll]: number of followers of the target account;
\item [$t_{i}$]: number of tweets of the \textit{i}-th follower of the
  target account;
\item [$\fri_{i}$]: number of friends of the \textit{i}-th follower of
  the target account;
\item [$\foll_{i}$]: number of followers of the \textit{i}-th follower
  of the target account.
\end{enumerate}
\begin{table}
\footnotesize
  \centering
  \begin{tabular}{lccc}
    \toprule
    ~ & \textbf{profile} & \textbf{timeline} & \textbf{relationships} \\
    \midrule
    API calls &
    $\left\lceil\frac{\foll}{100}\right\rceil$ &
    $\sum_{i \in \foll} \left\lceil\frac{t_i}{200}\right\rceil$ &
    $\sum_{i \in \foll} \left(\left\lceil\frac{\foll_i}{5000}\right\rceil+\left\lceil\frac{\fri_i}{5000}\right\rceil\right)$ \\[1ex]
    Best-case & $\left\lceil\frac{\foll}{100}\right\rceil$ & $\foll$ & $2*\foll$ \\[1ex]
    Worst-case & $\left\lceil\frac{\foll}{100}\right\rceil$ & $16*\foll$ & unpredictable \\[1ex]
    Calls/min. & $12$ & $12$ & $1$ \\
    \bottomrule
  \end{tabular}
  \caption {Number of API calls needed to download data.} 
  \label{tab:singleAbs}
\end{table}

The number of API calls for each category depends on the maximum
number of accounts (100), tweets (200) and friends/followers (5000)
that can be fetched from Twitter with a single API request. For
example, for the \textit{profile} category, a single API call can
return 100 follower profiles, leading to
$\left\lceil\frac{\foll}{100}\right\rceil$ API calls in total.  The detailed
costs do not account for the initial download of the whole list of
\foll\ followers of the target account, that requires
$\left\lceil\frac{\foll}{5000}\right\rceil$ API calls.

Table~\ref{tab:singleAbs} also shows the minimum (\textit{Best-case})
and maximum (\textit{Worst-case}) number of API calls that could
possibly be required, that depend on the length of the timelines and
the number of relationships of the followers. The Best-case is when
one single API call is sufficient to get all the data for a single
follower. For the Worst-case we can only precisely evaluate the number
of API calls for the \textit{timeline} category, since the number of
tweets that can be accessed from a user timeline is limited to 3200,
leading to a maximum of 16 calls for each follower. The number of
friends and followers, instead, is not limited and, therefore, it is
impossible to calculate a worst-case scenario for the
\textit{relationship} category. However, just to provide an
estimation, we can consider the account with the highest number of
followers on Twitter, which, at the time of writing, belongs to the
pop star Katy Perry (\textit{@katyperry}), with about 60 millions of
followers. We can therefore consider as the worst-case scenario an
account with 60 millions followers and 60 millions friends, which
leads to a number of API calls equal to $22000*\foll$.

Observing the values of Table~\ref{tab:singleAbs}, we have a clear
idea of the order of magnitude of each class: features in \textit{Class
  B}  are 100 times more costly than features of \textit{Class A}, 
while features of \textit{Class C} could be several orders
 of magnitude more costly than features of \textit{Class A}.

 Furthermore, to protect Twitter from abuse, the number of API calls
 allowed per minute is limited.  In Table~\ref{tab:singleAbs}, we also
 report the maximum number of calls allowed per minute
 (\textit{Calls/min.}), which directly impacts on the time needed to
 complete the data acquisition.

Some further considerations
 follow. Firstly, data
collected for a category can be used to evaluate all the features of
that category. Secondly, Twitter limits the number of calls of the same API, but different APIs can be called in parallel. 
This means that data for all the three feature categories
can be possibly acquired concurrently. The total time required
to collect all 
 data depends on the category that requires more
time, i.e., the \textit{relationship} one. In other words, to get the total time, one should not consider the sum of  the time needed for each of the three cost classes, but  just the most costly one.

\subsection{The \textit{Class A} classifier}\label{sec:text-a-class}
\makeatletter{}\begin{table}[t!]
\footnotesize
\centering
\begin{tabular}{llrrrrrr}
\toprule
&&\multicolumn{6}{c}{\textbf{evaluation metrics}}\\
\cmidrule{3-8}
&\textbf{algorithm} & \textit{accuracy} &\textit{precision}&\textit{recall}& \textit{F-M.} & \textit{MCC} & \textit{AUC}\\
\midrule
\multicolumn{8}{l}{\textit{Class C classifiers that use all the features}}\\
RF&Random Forest&  \textbf{0.994}  &\textbf{0.997}  &0.990  &\textbf{0.994}  &\textbf{0.987}  &\textbf{0.999}  \\
D&Decorate	& 0.993  &0.993  &\textbf{0.993}  &0.993  &\textbf{0.987}  &\textbf{0.999}  \\
J48&Decision Tree&  0.992  &0.991  &0.992  &0.992  &0.983  &0.993  \\
AB&Adaptive Boosting	&  0.987  &0.988  &0.987  &0.987  &0.975  &\textbf{0.999}  \\
BN&Bayesian Network	&  0.960  &0.965  &0.954  &0.960  &0.921  &0.991  \\
kNN&k-Nearest Neighbors	  &0.971 &0.963  &0.979  &0.971  &0.941  &0.990\\
LR&Logistic Regression	  &0.986  &0.985  &0.988  &0.986  &0.972  &0.996\\
SVM&Support Vector Machine	  &0.987  &0.983  &0.991  &0.987  &0.974  &0.987\\
\midrule
\multicolumn{8}{l}{\textit{Class A classifiers that use only Class A (profile) features}}\\
RF&Random Forest&  \textbf{0.987}  &\textbf{0.993}  &0.980  &\textbf{0.987}  &\textbf{0.967}  &\textbf{0.995}  \\
D&Decorate	&  0.984  &0.987  &\textbf{0.981}  &0.984  &0.964  &\textbf{0.995}  \\
J48&Decision Tree& 0.983  &0.987  &0.979  &0.983  &0.962  &0.983  \\
AB&Adaptive Boosting	&  0.972  &0.975  &0.969  &0.972  &0.941  &\textbf{0.995}  \\
BN&Bayesian Network	&  0.966  &0.969  &0.963  &0.966  &0.928  &0.991  \\
kNN&k-Nearest Neighbors	  &0.957 &0.961  &0.953  &0.957  &0.914  &0.978\\
LR&Logistic Regression	  &0.971  &0.964  &0.978  &0.971  &0.942  &0.987\\
SVM&Support Vector Machine	  &0.961  &0.972  &0.950  &0.961  &0.923  &0.961\\
\bottomrule
\end{tabular}
\caption{Performance comparison for 10-fold cross validation. Training set: \set{BAS}.}
\label{table:classifiers}
\end{table}

All the rules and features considered in this study fall into one of
the three aforementioned categories, as reported in
Table~\ref{tab:feature-categories}.  Therefore, their crawling cost
impacts on the final cost of the whole feature set and, ultimately, to
the class of the classifier: a classifier that uses a certain feature
set belongs to the class of the more expensive feature. Then, all the
classifiers of the previous sections are classifiers of \textit{Class
  B}, with the exception of the classifier with the feature set of
Yang \textit{et al.}~\cite{Yang:2013}, that belongs to \textit{Class
  C}.

In the following, we consider a lightweight classifier working only
with features of \textit{Class A}. We aim at verifying whether the
\textit{Class A} classifier reaches performances that are comparable to those of the 
most expensive \textit{Class B} and \textit{Class C} classifiers.

Table~\ref{table:classifiers} reports the results of the classifiers built on our \emph{BAS}\ dataset, using two different feature sets: all the features (independently from their cost) and the \textit{Class A} features.
We start observing that the \textit{Class C} classifiers, built over all the features considered in our study, perform better than all the other ones, including the classifiers using the feature sets of Yang \textit{et al.}~\cite{Yang:2013} and Stringhini \textit{et al.}~\cite{Stringhini:2010}, as reported in~Table~\ref{table:classifiersSezione5}.
In particular, RF, D and AB achieve an \textit{AUC} as high as 0.999.
 \textit{Class C} classifier only slightly outperforms \textit{Class A} classifier. Indeed, there is a difference of around 0.02 in \textit{MCC} for RF, D and J48.
The \textit{AUC} reduction is even smaller, only 0.004 for RF, D, J48 and 0 for BN.
It is worth noting that the \textit{Class A} classifier with BN outperforms the \textit{Class C} competitor, noticeably obtaining an increase in all the metrics, but the \textit{AUC}.
Instead, the \textit{Class A} kNN, LR and SVM classifiers suffer from a significant drop in performances with respect to the \textit{Class C} counterparts.

Concerning the complexity of the two decision trees obtained with the J48 algorithm, they are comparable, since both of them are composed by 31 nodes and 16 leaves and they have a height of 7.
A further interesting observation is that both the classifiers employ a set of features that includes both features proposed by the grey literature for fake detection, which we introduced in Section~\ref{sec:fakedet}, and by Academia for spam and bot detection, as detailed in Section~\ref{sec:spamdet}.

  The analysis carried out in Section~\ref{sec:costs} highlights the
  effort and time needed to compute many of the features commonly
  proposed for the detection of spam and fake accounts. As shown by
  the Katy Perry example, crawling costs for some of the proposed
  features are totally infeasible for accounts with hundreds of
  thousands or millions of followers.  For this reason, we trained and
  evaluated the proposed \textit{Class A} classifiers, which achieve
  overall good performances, while only exploiting cost-efficient
  features.  The \textit{Class A} classifiers thus represent a
  feasible solution for the investigation of fake followers on a large
  scale.

  We have to point out, however, that countermeasures could be taken
  to evade some of the simplest features our classifiers are built
  upon~\cite{Yang:2013}. This would require to continually
  monitor and update the choice of such features, to keep pace with
  the fake follower generators. While contemporary fake Twitter
  followers are effectively and efficiently spotted by our
  \textit{Class A} classifier, we can consider the use of the most
  expensive \textit{Class B} or even \textit{Class C} features to have
  stronger evidences about the more suspicious followers.

\subsection{Validation of the \textit{Class A} classifier}
\label{sec:valid-text-a}
 In this section, we propose a
  validation of our \textit{Class A} classifier, built with our
  baseline \emph{BAS} dataset. In particular, we set up two different
  experiments based on a random sampling of Twitter accounts.
For the first experiment, we built a set of 1000
Twitter accounts, randomly selecting numeric Twitter user IDs, ranging
from user ID 12 (the very first valid Twitter account -- \textit{@jack} -- belonging to Jack Dorsey,
founder of Twitter) to the user ID representing the last Twitter
account created at the time of our experiment.  This represents an
unbiased sample of all the currently available (i.e., not closed,
banned or suspended) Twitter accounts. This test set therefore
comprises a broad range of accounts created during the nine years
since Twitter's advent.
For the second experiment, instead, we
consider a random sample of 1500 accounts among the followers
of the US President Barack Obama -- \textit{@BarackObama}.  This experiment
resembles the practical application scenario of the proposed
classifier, since it investigates a sample of a single account's
followers, in this case a major politician.

All the accounts
acquired with the two aforementioned approaches have been labeled as
humans, following the same approach used in~\cite{Stringhini:2013}.
Twitter officially reports that fake and spam profiles together are less than 5\% of all registered accounts~\cite{twitters1form}, therefore we are confident that just a few among the accounts labeled as humans might actually be fake ones. Automatically labeling the sampled accounts as humans would result in an error of at most 5\%,  still allowing an overall correct validation of our classifiers. In addition, many of the accounts randomly acquired for the first and second experiment show few signs of activity on Twitter (more than 70\% of them did not post a tweet in the 3 months prior to our data acquisition). Thus, more reliable checks like CAPTCHA-based verifications would result in very sparse answers. Furthermore, we believe that also including less active accounts in our test sets, allows to validate our classifiers with Twitter accounts having different characteristics than those of our baseline dataset of humans in Section~\ref{subsec:humanset} (e.g., showing fewer ``human" features).
Together with the human accounts, the
test set  also includes the 1401 fake followers we bought, but not
 included in the \emph{BAS} dataset that we used as training set (Section~\ref{subsec:baseset}).

\makeatletter{}\begin{table}[t!]
\footnotesize
\centering
\begin{tabular}{llrrrrrr}
\toprule
&&\multicolumn{6}{c}{\textbf{evaluation metrics}}\\
\cmidrule{3-8}
&\textbf{algorithm} & \textit{accuracy} &\hspace{-8pt}\textit{precision}&\textit{recall}& \textit{F-M.} & \textit{MCC} & \textit{AUC}\\
\midrule
\multicolumn{8}{l}{\textit{Class A validation on a test set of random sampled accounts}}\\
RF&Random Forest  &\textbf{0.975}  &\textbf{0.982}  &\textbf{0.975}  &\textbf{0.979}  &\textbf{0.949}  &\textbf{0.989} \\
D&Decorate       &0.904  &0.894  &0.948  &0.920  &0.802  &0.975 \\
J48&Decision Tree&0.904  &0.898  &0.942  &0.920  &0.801  &0.962 \\
AB&Adaptive Boosting    &0.767  &0.737  &0.936  &0.825  &0.526  &0.959 \\
BN&Bayesian Network      &0.891  &0.876  &0.947  &0.910  &0.776  &0.961 \\
kNN&k-Nearest Neighbors	  &0.946 &0.962  &0.944  &0.953  &0.889  &0.969\\
LR&Logistic Regression	  &0.551  &0.922  &0.252  &0.396  &0.299  &0.827\\
SVM&Support Vector Machine	  &0.955  &\textbf{0.982}  &0.941  &0.961  &0.910  &0.958\\
\midrule
\multicolumn{8}{l}{\textit{Class A validation on a test set of random sampled Obama followers}}\\
RF&Random Forest  & \textbf{0.929}  &\textbf{0.889}  &\textbf{0.975}  &\textbf{0.930}  &\textbf{0.862}  &\textbf{0.970} \\
D&Decorate       & 0.909  &0.875  &0.948  &0.910  &0.820  &0.964 \\
J48&Decision Tree& 0.902  &0.868  &0.942  &0.903  &0.807  &0.924 \\
AB&Adaptive Boosting    & 0.862  &0.810  &0.936  &0.868  &0.733  &0.949 \\
BN&Bayesian Network      & 0.786  &0.710  &0.947  &0.811  &0.607  &0.943 \\
kNN&k-Nearest Neighbors	  &0.733 &0.763  &0.655  &0.705  &0.469  &0.828\\
LR&Logistic Regression	  &0.615  &0.784  &0.290  &0.423  &0.278  &0.794\\
SVM&Support Vector Machine	  &0.873  &0.851  &0.897  &0.873  &0.748  &0.874\\
\bottomrule
\end{tabular}
\caption{Class A classifier validation on two different test sets.}
\label{table:classa:validation}
\end{table}

In Table~\ref{table:classa:validation}, we report the results of the experiments on the test sets.
Comparing results of Table~\ref{table:classa:validation} with those of Tables~\ref{table:classifiersSezione5} and~\ref{table:classifiers}, we notice a broader range of performances. Indeed, while in Tables~\ref{table:classifiersSezione5} and~\ref{table:classifiers} almost all the considered classifiers achieved comparable performances, here we can see some major differences. This means that the increased difficulty of the detection task on the two considered test sets highlighted differences that were not visible in the previous  experiments.

From Table~\ref{table:classa:validation}, we can observe that the validation against sampled Obama followers (bottom half of the table) proves to be more error prone than that against random sampled accounts.
Indeed, almost all the classifiers achieve worse results in the former experiment, as deducible, for instance, from the \textit{AUC} scores.

Observing the validation against the random sampled accounts, we can see that the five \textit{Class A} classifiers obtain an accuracy above 0.9, a precision close to (when not much higher than) 0.9 and a recall above 0.94: this means that they are able to spot almost all the fake followers of the test set.
This is particularly true for the best performing classifier, RF, that reaches 0.975 for both accuracy and recall, with a precision of 0.982.
The highest performances are also shown by both F-measure and \textit{MCC} values, that are noticeably higher for RF when compared with the others.
AB and BN obtain lower results for both accuracy and precision, but still a noticeably high recall, meaning that only few fake followers are left behind, but a considerably higher number of human accounts are classified as fake followers. Among all the classifiers, LR is the one achieving the worst performances, because of the low \textit{MCC} values obtained with both the test sets: 0.299 and 0.278 respectively. For the validation against sampled Obama followers, also kNN achieves unsatisfactory performances, with \textit{MCC} equal to 0.469. The SVM classifier, instead, proves to be extremely effective, being the second best classifier when validated against random sampled accounts, and still performing very well against sampled Obama followers.

The RF, D, and J48 classifiers obtain results that are comparable for the two test sets, with an accuracy greater than 0.9 and a precision close to 0.87, with a reduction of only 0.025 on average.
This is also confirmed by both the F-measure and \textit{MCC} that are very close to the results above. In the two validation experiments, AB and BN noticeably switch their performances: the Adaptive Boosting algorithm raises its accuracy of 0.1 and its precision of around 0.08, outperforming the Bayesian Network-based one. This latter looses 0.105 of accuracy, meaning that many of the accounts we labeled as humans were recognized as fake followers.
Also the RF classifier looses 0.4 of accuracy and a considerable 0.1 in precision, meaning that it considers as fake followers many of the sampled Obama's followers.
We can finally observe that the \textit{AUC} metric is always very high and that does not consistently reflect the real performances obtained by the five classifiers, as the F-measure and the \textit{MCC} actually do.

\subsection{Assessing the global importance of \textit{Class A} features}
\label{sec:feature-relevance}
Motivated by the results obtained by our \textit{Class A} classifiers, we went further to asses the global importance of \textit{Class A} features towards the detection of fake Twitter followers. In order to estimate the importance of the single features among all the 8 classifiers, we followed the information fusion-based sensitivity analysis technique adopted in~\cite{sevim2014developing}. Information fusion is a technique aimed at leveraging the predictive power of several different models in order to achieve a combined prediction accuracy which is better than the predictions of the single models~\cite{chase2000composite}. Sensitivity analysis, instead, aims at assessing the relative importance of the different features used to build a classification model~\cite{saltelli2002making}. It is therefore possible to combine information fusion and sensitivity analysis to estimate the global importance of several features, used in different machine learning classifiers, towards a common classification task.

\makeatletter{}\begin{table}[t!]
\footnotesize
\centering
\begin{tabular}{clrrl}
\toprule
\textbf{rank} & \textbf{feature} & \textbf{proposed in} & \multicolumn{2}{l}{\textbf{normalized score}} \\
\midrule
1 & friends/(followers\string^2) ratio & \cite{Stringhini:2010} & 1.000 & \mybar{1.000}\\
2 & age & \cite{Yang:2013, Lee:2010, miller2014twitter, chu2012detecting}  & 0.919 & \mybar{0.919}\\
3 & number of tweets & \cite{chu2012detecting, miller2014twitter, Stringhini:2010, Camisani:2012, SocialBakers}  & 0.816 & \mybar{0.816}\\
4 & profile has name & \cite{Camisani:2012}  & 0.782 & \mybar{0.782}\\
5 & number of friends & \cite{bhat2014using, Stringhini:2010, SocialBakers, miller2014twitter}  & 0.781 & \mybar{0.781}\\
6 & has URL in profile & \cite{Camisani:2012}  & 0.768 & \mybar{0.768}\\
7 & following rate & \cite{Yang:2013}  & 0.765 & \mybar{0.765}\\
8 & default image after 2 months & \cite{SocialBakers}  & 0.755 & \mybar{0.755}\\
9 & belongs to a list & \cite{Camisani:2012}  & 0.752 &\mybar{0.752} \\
10 & profile has image & \cite{Camisani:2012}  & 0.751 & \mybar{0.751}\\
11 & friends/followers $\ge$ 50 & \cite{SocialBakers}  & 0.736 & \mybar{0.736}\\
12 & \textit{bot} in biography & \cite{blogger1:2012}  & 0.734 & \mybar{0.734}\\
13 & duplicate profile pictures & \cite{blogger1:2012}  & 0.731 & \mybar{0.731}\\
14 & 2 $\times$ followers $\ge$ friends & \cite{blogger1:2012}  & 0.721 & \mybar{0.721}\\
15 & friends/followers $\simeq$ 100 & \cite{blogger1:2012}  & 0.707 & \mybar{0.707}\\
16 & has address & \cite{Camisani:2012}  & 0.677 & \mybar{0.677}\\
17 & no bio, no location, friends $\ge$ 100 & \cite{SocialBakers}  & 0.664 & \mybar{0.664}\\
18 & has biography & \cite{Camisani:2012}  & 0.602 & \mybar{0.602}\\
19 & number of followers & \cite{Camisani:2012, miller2014twitter, ahmed2013generic}  & 0.594 & \mybar{0.594}\\
\bottomrule
\end{tabular}
\caption{Global importance of \textit{Class A} features for the detection of fake Twitter followers.}
\label{table:classa-feat-rel}
\end{table}

To reach this goal, we firstly re-train each of the 8 \textit{Class A} classifiers with our baseline \emph{BAS} dataset, by removing one feature at a time. This leads to $19 \times 8$ different classifiers. Subsequently, each of those classifiers is evaluated against our test sets. The \textit{local} sensitivity score for the $i$-th feature of the $j$-th classifier can be computed as $S_{j,i} = \frac{MCC_{j,-i}}{MCC_{j}}$, namely as the reduction in classification performance of $j$ when used without the feature $i$ with respect to the classification performance of $j$ using all the 19 \textit{Class A} features.
Then, we can compute a \textit{global} sensitivity score for the $i$-th feature via a weighted sum of the local sensitivities, $S_i = \sum\limits_{j}w_jS_{j,i}$. The weighting factor $w_j$ is proportional to the \textit{MCC} of the $j$-th classifier, as reported in Table~\ref{table:classa:validation}, so that the best performing classifiers are weighted more. Finally, we rank features and assess their relative importance by normalizing their global sensitivities, so that the best performing feature has a normalized score of 1.

Table~\ref{table:classa-feat-rel} shows the results of this information fusion-based sensitivity analysis. Notably, none of the features is clearly dominant: all the 19 \textit{Class A} features give a significant contribution to the detection of fake Twitter followers. Results of this in-depth analysis also show a considerable agreement with our feature evaluation studies reported in Tables~\ref{table:results} and~\ref{table:feature:set}. Specifically, the most important feature of Table~\ref{table:classa-feat-rel}, namely \textit{$friends/(followers\string^2)$ ratio}, is the second best feature in Table~\ref{table:feature:set}, outperformed only by the \textit{Class C} feature \textit{bidirectional link ratio}. Given the relevant contribution of other features such as \textit{number of friends} and \textit{following rate}, we believe that features aimed at evaluating the social ties between an account and its neighbors play a central role towards the detection of fake Twitter followers.

\makeatletter{}\section{Conclusions}
\label{sec:conc}
In this paper, we focused on efficient techniques for fake Twitter followers detection. 

To reach the goal, we firstly created a baseline dataset of human and
fake follower accounts, the latter being bought from available online
markets. Then, we surveyed various proposals for spammer and bot
detection, based on classification rules and feature sets. Such
proposals come partly from Academia and partly from the grey literature.
Such rules and features were eventually tested on our dataset to understand their
effectiveness in detecting fake Twitter followers.  A few features
were selected and used by a set of classifiers that we have trained on
the baseline dataset. 
Going further, we ranked the best performing features 
according to their crawling cost.  This led us to identify three
categories of features belonging to three different, increasing, cost classes. Finally,
we built a series of classifiers 
that only leverage cost-effective (\textit{Class A}) features. With this final outcome, we were able 
 to achieve detection rates comparable with the best of breed
classifiers, whereas these latter necessitate overhead-demanding
features.

  Among the results of this study, there is the construction of a baseline dataset of verified human and fake follower accounts. To foster research on the novel issue of fake Twitter followers, we have publicly released our dataset to the scientific community~\cite{fakeproject}.

For our analysis, we considered 49 distinct features and 8 different ``glass-box'' and ``black-box" machine learning classifiers. So-called ``glass-box" classifiers produce interpretable models and an analysis of their inner structure allows to get insights into the mechanisms exploited to perform the detection. In turn, this is useful  to better understand the characteristics of Twitter accounts that can be leveraged to discriminate between fake and genuine followers.
Nonetheless, ``black-box" classifiers have been recently employed in a broad range of diverse classification tasks, achieving excellent results. Therefore, for the sake of experimentation, we expanded the set of machine learning algorithms employed in previous works on fake and spam detection, with the adoption of powerful classifiers, such as Support Vector Machines (SVM). Indeed, the results of our analyses confirm that SVMs achieve results comparable to those of the best performing classifiers (such as Random Forest and Decorate).
The adoption of other powerful classification techniques, such as Artificial Neural Networks (ANNs), could provide other interesting results and, therefore, we consider its adoption and evaluation promising ground for future work.

Among all the analyzed features, we have seen that those yielding
  the best results are the ones based on the friends and followers of
  the account under investigation, such as the \textit{bidirectional
    link ratio} and the \textit{$friends/(followers\string^2)$ ratio}.
  By evaluating the social ties between an account and its neighbors,
  these features are more effective than those based on simpler
  characteristics, like, e.g., other information in the account's
  profile. However, relationships-based (\textit{Class C}) features are more demanding in
  terms of data to be downloaded and, consequently, they require a significant  analysis time,
  making them unsuitable for analyses on massive amounts of followers.   Timeline-based (\textit{Class B}) features have been shown to be less time-demanding, while still effective; hence, these
   might represent a promising trade-off between efficient and
  accurate detections. Anyway, as shown by the proposed \textit{Class
    A} classifiers, the detection of currently available fake Twitter
  followers is possible even without leveraging resource-demanding features, by means
  of efficient algorithms and accurate selection and combination of
  less-demanding features. We also  evaluated the global importance of the \textit{Class A} features for the fake Twitter follower detection, using information fusion-based sensitivity analysis and showing that those based on social
relations play a dominant role.

As future work, we aim at designing and testing other advanced features that could be added to our lightweight fake
  followers' classifier, leveraging additional characteristics of Twitter accounts. Data that could be further exploited for the classification task are the content of tweets and
  the accounts' behavior. In particular, bot development forums represent a fruitful source of information to know more about bot/fake/spam accounts design, in terms, e.g., of similarities and differences in their behavior. 
  As highlighted by our work, the difficulty of the detection task and the massive numbers of accounts to analyze ask for the adoption of features that are not only \textit{effective}, but also \textit{efficient}, with regard to their crawling costs.
Therefore, we believe that future works along this line of research should consider the balance between the predictive power of new features and their cost. This would allow to improve the detection of fake Twitter followers, while still retaining a scalable approach, making \textit{en masse} analysis practically feasible.

It is foreseeable that countermeasures will be taken to masquerade
  a fake account with respect to some of the simplest features which
  our classifiers are built upon. This would require to continually
  monitor and update the choice of such features, conceivably studying
  (or directly interacting with) the same generators of fakes, to keep their
  pace.
  Therefore, we believe that the features renovation process can be considered 
  as another interesting direction    for future research.

\makeatletter{}\section*{Acknowledgements}
\begin{small}
The authors would like to thank the anonymous reviewers who helped improving the quality of the manuscript.
\end{small} 
\section*{\refname}
\bibliographystyle{elsarticle-num}
\bibliography{references}

\begin{thebibliography}{10}
\expandafter\ifx\csname url\endcsname\relax
  \def\url#1{\texttt{#1}}\fi
\expandafter\ifx\csname urlprefix\endcsname\relax\def\urlprefix{URL }\fi
\expandafter\ifx\csname href\endcsname\relax
  \def\href#1#2{#2} \def\path#1{#1}\fi

\bibitem{chu2012detecting}
Z.~Chu, I.~Widjaja, H.~Wang, Detecting social spam campaigns on {Twitter}, in:
  Applied Cryptography and Network Security, Springer, 2012, pp. 455--472.

\bibitem{Yang:2013}
C.~Yang, R.~Harkreader, G.~Gu, Empirical evaluation and new design for fighting
  evolving {Twitter} spammers, Information Forensics and Security, IEEE
  Transactions on 8~(8) (2013) 1280--1293.

\bibitem{ahmed2013generic}
F.~Ahmed, M.~Abulaish, A generic statistical approach for spam detection in
  online social networks, Computer Communications 36~(10) (2013) 1120--1129.

\bibitem{miller2014twitter}
Z.~Miller, B.~Dickinson, W.~Deitrick, W.~Hu, A.~H. Wang, Twitter spammer
  detection using data stream clustering, Information Sciences 260 (2014)
  64--73.

\bibitem{cha2010measuring}
M.~Cha, H.~Haddadi, F.~Benevenuto, P.~K. Gummadi, Measuring user influence in
  {Twitter}: The million follower fallacy, ICWSM 10~(10-17) (2010) 30.

\bibitem{castillo2011information}
C.~Castillo, M.~Mendoza, B.~Poblete, Information credibility on {Twitter}, in:
  Proceedings of the 20th international conference on World Wide Web, ACM,
  2011, pp. 675--684.

\bibitem{asonam14}
J.~S. Alowibdi, U.~A. Buy, P.~S. Yu, L.~Stenneth, Detecting deception in online
  social networks, in: Advances in Social Networks Analysis and Mining
  (ASONAM), 2014 IEEE/ACM International Conference on, 2014, pp. 383--390.

\bibitem{Stringhini:2010}
G.~Stringhini, C.~Kruegel, G.~Vigna, Detecting spammers on social networks, in:
  26th Annual Computer Security Applications Conference, ACSAC '10, ACM, 2010,
  pp. 1--9.

\bibitem{SocialBot11}
Y.~Boshmaf, I.~Muslukhov, K.~Beznosov, M.~Ripeanu, The socialbot network: When
  bots socialize for fame and money, in: 27th Annual Computer Security
  Applications Conference, ACSAC '11, ACM, New York, NY, USA, 2011, pp.
  93--102.

\bibitem{ChuGWJ:2012}
Z.~Chu, S.~Gianvecchio, H.~Wang, S.~Jajodia, Detecting automation of {Twitter}
  accounts: Are you a human, bot, or cyborg?, IEEE Trans. Dependable Sec.
  Comput. 9~(6) (2012) 811--824.

\bibitem{blogger1:2012}
{Stateofsearch.com}, {How to recognize Twitterbots: 7 signals to look out
  for.}, http://goo.gl/YZbVf, Last checked 23/07/15 (August 2012).

\bibitem{Camisani:2012}
{Camisani-Calzolari M. (digitalevaluations)}, {Analysis of Twitter followers of
  the US Presidential Election candidates: Barack Obama and Mitt Romney}
  (August 2012).

\bibitem{StatusPeople}
{Statuspeople.com}, {Status People Fakers}, http://goo.gl/0Jpky, Last checked
  23/07/15.

\bibitem{SocialBakers}
SocialBakers, Fake follower check, http://goo.gl/chWn0, Last checked 23/07/15.

\bibitem{TwitterAudit}
{Twitteraudit}, {How many of your followers are real?}, www.twitteraudit.com,
  Last checked 23/07/15.

\bibitem{DASec:2014}
S.~Cresci, R.~{Di Pietro}, M.~Petrocchi, A.~Spognardi, M.~Tesconi, {A Criticism
  to Society (as seen by Twitter analytics)}, in: 1st International Workshop on
  Big Data Analytics for Security, IEEE, 2014, pp. 194--200.

\bibitem{GaoCLPC12}
H.~Gao, Y.~Chen, K.~Lee, D.~Palsetia, A.~N. Choudhary, Towards online spam
  filtering in social networks, in: 19th Annual Network and Distributed System
  Security Symposium, {NDSS} 2012, San Diego, California, USA, February 5-8,
  2012, 2012.

\bibitem{Lee:2013}
S.~Lee, J.~Kim, {WarningBird}: A near real-time detection system for suspicious
  {URLs} in {Twitter} stream, IEEE Trans. Dependable Secur. Comput. 10~(3)
  (2013) 183--195.

\bibitem{ThomasGMPS11}
K.~Thomas, C.~Grier, J.~Ma, V.~Paxson, D.~Song, Design and evaluation of a
  real-time {URL} spam filtering service, in: 32nd {IEEE} Symposium on Security
  and Privacy, S{\&}P 2011, 22-25 May 2011, Berkeley, California, {USA}, 2011,
  pp. 447--462.

\bibitem{zubiaga2014tweet}
A.~Zubiaga, H.~Ji, Tweet, but verify: epistemic study of information
  verification on {Twitter}, Social Network Analysis and Mining 4~(1) (2014)
  1--12.

\bibitem{Gao14}
H.~Gao, Y.~Yang, K.~Bu, Y.~Chen, D.~Downey, K.~Lee, A.~Choudhary, Spam ain't as
  diverse as it seems: Throttling {OSN} spam with templates underneath, in:
  Annual Computer Security Applications Conference, ACSAC, 2014, pp. 76--85.

\bibitem{weibo14}
Y.~Liu, B.~Wu, B.~Wang, G.~Li, {SDHM}: A hybrid model for spammer detection in
  {Weibo}, in: Advances in Social Networks Analysis and Mining (ASONAM), 2014
  IEEE/ACM International Conference on, 2014, pp. 942--947.

\bibitem{data_mining}
M.~Hall, I.~Witten, E.~Frank, Data Mining: Practical Machine Learning Tools and
  Techniques, 3rd Edition, Morgan Kaufmann, 2011.

\bibitem{ComCom14}
S.~Lee, J.~Kim, Early filtering of ephemeral malicious accounts on {Twitter},
  Computer Communications 54 (2014) 48--57.

\bibitem{Stringhini:2012}
G.~Stringhini, M.~Egele, C.~Kruegel, G.~Vigna, {Poultry markets: on the
  underground economy of Twitter followers}, in: Workshop on online social
  networks, WOSN '12, ACM, 2012, pp. 1--6.

\bibitem{Stringhini:2013}
G.~Stringhini, G.~Wang, M.~Egele, C.~Kruegel, G.~Vigna, H.~Zheng, B.~Y. Zhao,
  {Follow the Green: Growth and Dynamics in Twitter Follower Markets}, in:
  Internet Measurement Conference, IMC '13, 2013, pp. 163--176.

\bibitem{Thomas2013}
K.~Thomas, D.~McCoy, C.~Grier, A.~Kolcz, V.~Paxson, Trafficking fraudulent
  accounts: The role of the underground market in {Twitter} spam and abuse, in:
  22nd USENIX Security Symposium, USENIX, Washington, D.C., 2013, pp. 195--210.

\bibitem{Lee:2010}
K.~Lee, J.~Caverlee, S.~Webb, Uncovering social spammers: social honeypots +
  machine learning, in: SIGIR Conference on Research and Development in
  Information Retrieval, ACM, 2010, pp. 435--442.

\bibitem{Yardi}
S.~Yardi, D.~M. Romero, G.~Schoenebeck, D.~Boyd, {Detecting Spam in a Twitter
  Network}, First Monday 15~(1).

\bibitem{yan2013peri}
G.~Yan, {Peri-Watchdog}: Hunting for hidden botnets in the periphery of online
  social networks, Computer Networks 57~(2) (2013) 540--555.

\bibitem{mccord:2011}
M.~McCord, M.~Chuah, Spam detection on {Twitter} using traditional classifiers,
  in: Autonomic and Trusted Computing, Vol. 6906 of LNCS, Springer Berlin
  Heidelberg, 2011, pp. 175--186.

\bibitem{bhat2014using}
S.~Y. Bhat, M.~Abulaish, Using communities against deception in online social
  networks, Computer Fraud \& Security 2014~(2) (2014) 8--16.

\bibitem{fakeproject}
{The Fake Project}, Dataset, http://mib.projects.iit.cnr.it/dataset.html.

\bibitem{weiss2003learning}
G.~M. Weiss, F.~Provost, Learning when training data are costly: the effect of
  class distribution on tree induction, Journal of Artificial Intelligence
  Research 19 (2003) 315--354.

\bibitem{twitters1form}
{Twitter Inc.}, {Twitter's IPO filing},
  http://www.sec.gov/Archives/edgar/data/1418091/000119312513390321/d564001ds1.htm,
  Last checked 23/07/15 (Oct 2013).

\bibitem{kohavi98}
R.~Kohavi, F.~Provost, Glossary of terms, Machine Learning 30~(2-3) (1998)
  271--274.

\bibitem{Baldi2000}
P.~Baldi, S.~Brunak, Y.~Chauvin, C.~Andersen, H.~Nielsen, {Assessing the
  accuracy of prediction algorithms for classification: an overview},
  Bioinformatics 16~(5) (2000) 412--424.

\bibitem{powers2011evaluation}
D.~M.~W. Powers, Evaluation: from precision, recall and {F}-measure to {ROC},
  informedness, markedness and correlation, International Journal of Machine
  Learning Technologies 2~(1) (2011) 37--63.

\bibitem{friedman2001elements}
J.~Friedman, T.~Hastie, R.~Tibshirani, The elements of statistical learning,
  Vol.~1, Springer Series in Statistics, 2001.

\bibitem{Mitchell:1997:ML:541177}
T.~M. Mitchell, Machine Learning, 1st Edition, McGraw-Hill, Inc., New York, NY,
  USA, 1997.

\bibitem{rice2006mathematical}
J.~Rice, Mathematical statistics and data analysis, Cengage Learning, 2006.

\bibitem{guyon2003introduction}
I.~Guyon, A.~Elisseeff, An introduction to variable and feature selection, The
  Journal of Machine Learning Research 3 (2003) 1157--1182.

\bibitem{weka}
M.~Hall, E.~Frank, G.~Holmes, B.~Pfahringer, P.~Reutemann, I.~H. Witten, The
  {WEKA} data mining software: an update, ACM SIGKDD explorations newsletter
  11~(1) (2009) 10--18.

\bibitem{chang2011libsvm}
C.-C. Chang, C.-J. Lin, {LIBSVM}: A library for {Support Vector Machines}, ACM
  Transactions on Intelligent Systems and Technology (TIST) 2~(3) (2011) 27.

\bibitem{hawkins2004problem}
D.~M. Hawkins, The problem of overfitting, Journal of chemical information and
  computer sciences 44~(1) (2004) 1--12.

\bibitem{sevim2014developing}
C.~Sevim, A.~Oztekin, O.~Bali, S.~Gumus, E.~Guresen, Developing an early
  warning system to predict currency crises, European Journal of Operational
  Research 237~(3) (2014) 1095--1104.

\bibitem{chase2000composite}
C.~W. Chase~Jr, Composite forecasting: combining forecasts for improved
  accuracy, Journal of Business Forecasting Methods \& Systems 19~(2) (2000) 2.

\bibitem{saltelli2002making}
A.~Saltelli, Making best use of model evaluations to compute sensitivity
  indices, Computer Physics Communications 145~(2) (2002) 280--297.

\end{thebibliography}
\newpage

\end{document}